\documentclass[11pt]{article}
\usepackage{amsthm,amsfonts,amsmath}
\swapnumbers
\theoremstyle{plain}

\newtheorem{thm}{THEOREM}[section]
\newtheorem{lm}[thm]{LEMMA}
\newtheorem{cl}[thm]{COROLLARY}

\theoremstyle{definition}

\theoremstyle{remark}
\newcommand{\upchi}{\raise1pt\hbox{$\chi$}}
\newcommand{\R}{{\mathord{\mathbb R}}}
\newcommand{\RR}{{\mathord{\mathbb R}}^3}
\newcommand{\C}{{\mathord{\mathbb C}}}

\newcommand{\an}{{\mathord{a^{\phantom{*}}_\lambda}}}

\newcommand{\cre}{{\mathord{a^*_\lambda}}}

\begin{document}

\title{\bf{Stability of a Model of Relativistic Quantum Electrodynamics}}
\author{\vspace{5pt} Elliott H. Lieb$^1$ and
Michael Loss$^{2}$ \\
\vspace{-4pt}\small{$1.$ Departments of Physics and Mathematics, Jadwin
Hall,} \\[-5pt]
\small{Princeton University, P.~O.~Box 708, Princeton, NJ
  08544}\\
\vspace{-4pt}\small{$2.$ School of Mathematics, Georgia Tech,
Atlanta, GA 30332}  \\ }
\date{September 2, 2001, revised March 15, 2002}
\maketitle

\footnotetext
[1]{Work partially
supported by U.S. National Science Foundation
grant PHY 98-20650-A02.}
\footnotetext
[2]{Work partially
supported by U.S. National Science Foundation
grant DMS 00-70589.\\
\copyright\, 2001 by the authors. This paper may be reproduced, in its
entirety, for non-commercial purposes.}

\begin{abstract} The relativistic ``no pair'' model of quantum
electrodynamics uses the Dirac operator, $D(A)$ for the electron
dynamics together with the usual self-energy of the quantized ultraviolet
cutoff electromagnetic field $A$ --- in the Coulomb gauge.  There are
no positrons because the electron wave functions are constrained to lie
in the positive spectral subspace of some Dirac operator, $D$, but the
model is defined for any number, $N$, of electrons, and hence describes a
true many-body system.  In addition to the electrons there are a number,
$K$, of fixed nuclei with charges $\leq Z$.  If the fields are not
quantized but are classical, it was shown earlier that such a model
is always unstable (the ground state energy $E=-\infty$) if one uses
the customary $D(0)$ to define the electron space, but is stable ($E >
-\mathrm{const.}(N+K)$) if one uses $D(A)$ itself  (provided the fine structure
constant $\alpha$ and $Z$ are not too large).  This result is extended
to quantized fields here, and stability is proved for $\alpha =1/137$
and $Z \leq 42$.  This formulation of QED is somewhat unusual because
it means that the electron Hilbert space is inextricably linked to the
photon Fock space.  But such a linkage appears to better describe the
real world of photons and electrons.

\end{abstract}
\section{Introduction} \label{intro}

The theory of the ground state of matter interacting with Coulomb forces
and with the magnetic field is not yet in a completely satisfactory state.
Open problems remain, such as the inclusion of relativistic mechanics
into the many-body formalism and the inclusion of the self-energy effects
of the radiation field, especially the quantized radiation field.

One of the fundamental attributes of quantum mechanics is the existence
of a Hamiltonian with a lowest, or ground state energy, and not merely
the existence of a critical point of a Lagrangian.  The `stability'
problem, which concerns us here, is to show that the ground state energy
is bounded below by a constant times the total number of particles, $N+K$,
where $N$ is the number of electrons and $K$ is the number of nuclei --
whose locations, in this model, are fixed, but chosen to minimize the energy. 
We do not discuss the
existence of a normalizable ground state eigenfunction, as in
\cite{GLL}, but only the lower boundedness of the Hamiltonian.

This problem has been resolved successfully in various models such as
the usual nonrelativistic Schr\"odinger Hamiltonian with only
electrostatic interactions.  Further developments include extensions to
relativistic kinetic energy $\sqrt{p^2+m^2}-m$ in place of the
nonrelativistic $p^2/2m$, and extensions to matter interacting with
classical magnetic fields (including a spin-field interaction $B$), 
stabilized by the classical field energy 
\begin{equation} \frac{1}{8
\pi} \int B(x)^2 {\rm d} x \ , \label{fieldenergy} 
\end{equation} 
and then the quantization of the $B$ field.  Many people
participated in this development and we refer the reader to \cite{LSS}
and the references therein for an account up to 1997.

In this paper we take a further step by addressing the problem of {\it
relativistic} matter, using the {\it Dirac operator} (without pair
production, i.e., the ``no-pair'' model) interacting with the {\it
quantized} radiation field having an ultraviolet cutoff $\Lambda$.  In
\cite{LSS} the corresponding problem was solved with a classical
radiation field, in which the field energy is given by
(\ref{fieldenergy}), and we shall use some of the ideas of that paper
here.  The idea for such a model goes back to \cite{BrownRavenhall1951}
and \cite{Sucher1980}.  With a classical $B$ field no ultraviolet cutoff
is needed, but it {\it is} needed with a quantized field, for otherwise
the field energy diverges.

Because of the ultraviolet cutoff our model, which in other respects is
relativistic, is not truly relativistic at energies of the order of the
cutoff. We have not, however,
attempted to renormalize the theory and, if this can be done consistently,
the resulting theory will be relativistic at all energies.

In \cite{BFG} the problem of {\it non}relativistic electrons (with spin)
interacting with the quantized ultraviolet cutoff field was solved by
using results in \cite{LLS} but using only the part of the field energy
within a distance $1/\Lambda $ of the fixed nuclei.  The constants and
exponents in \cite{BFG} were improved in \cite{FFG}; in particular,
the Hamiltonian is bounded below by $-\Lambda K$.  The relation of
the classical field energy to the quantized field energy involves a
commutator that, when integrated over the whole space $\R^3$ yields an
infinite constant, even with an ultraviolet cutoff.  This is the reason
for considering only a local field energy, since only a local field
energy yields a finite commutator, and we do the same here.

In Section \ref{basic} our model is defined and the main Theorem
\ref{mainthm} is stated.  With the fine structure constant $\alpha =
1/137$, stability holds for $Z\leq 42$.  The main idea of the ``no-pair''
model is that there are no positrons, and electronic wave functions
are allowed to lie only in the positive spectral subspace of some Dirac
operator $D$.  While the Dirac operator $D(A)$, which is contained in
the Hamiltonian and which defines the electron dynamics, always contains
the magnetic vector potential $A(x)$, the operator $D$ that defines
an electron could be $D(0)$, the free Dirac operator.  Indeed, this is
the conventional choice, but it is not gauge invariant and {\it always}
leads to instability as first shown in \cite{LSS} for classical fields
and here for quantized fields.

The question of instability is complicated. There are two kinds
(first and second) and two cases to consider (with and without
Coulomb potentials). Instability of the first kind means that the
ground state energy (bottom of  the spectrum of the  Hamiltonian)
is $-\infty$. Instability of the second kind means that the energy is
finite but is not bounded below by a constant times $N+K$. The occurence
of these instabilities may or may not depend on $\alpha$ and $Z$ and
whether or not a cutoff $\Lambda$ is present.

The physical nature of the instability, if it occurs, is different in
the two cases.  When it occurs in the absence of Coulomb potentials
(meaning that the $\alpha V_c$ term in (\ref{hamprime}) is omitted)
it is due to the $\sqrt{\alpha}\, A(x)$ term in  $D(A)$ blowing up.
When it occurs because of the Coulomb potentials being present it is
due to an electron falling into the Coulomb singularity of the nucleus.
The various possibilities, all proved in this paper, are summarized in
detail in the following two tables and discussed in detail in Appendix
\ref{sec:instability}.  For the proofs of the instabilities listed here,
we rely heavily on \cite{LSS} and \cite{GT}.

\bigskip
\centerline{\bf Electrons defined by projection onto the positive} 
\centerline{\bf subspace of $D(0)$, the free Dirac operator}

\bigskip
\quad\begin{tabular}{l||c|c|}
&Classical or quantized field  & Classical or quantized field \\
 &\quad without cutoff $\Lambda$ &  with cutoff $\Lambda$ \\
& $\alpha >0$ but arbitrarily small. & $\alpha >0$ but arbitrarily small.\\
 & & \\
\hline\hline
Without Coulomb&  Instability of &  Instability  of \\
potential $\alpha V_c$  & the first kind& the second kind \\
 \hline
With Coulomb &  Instability of &  Instability of \\
potential $\alpha V_c$ & the first kind &  the second kind \\
 \hline\hline
\end{tabular}

\bigskip\bigskip
\newpage

\centerline{\bf Electrons defined by projection onto the  positive}
\centerline{\bf subspace of $D(A)$, the Dirac operator with field}

\bigskip
\quad\qquad \qquad \begin{tabular}{l||c|c|}
&\multicolumn{2}{c| } {Classical field with or without cutoff $\Lambda$ } \\
&\multicolumn{2}{c| } {or quantized field with  cutoff $\Lambda$} \\
   &\multicolumn{2}{c | }   {}  \\
\hline\hline
Without Coulomb& \multicolumn{2}{c |} {The Hamiltonian is positive}  \\
potential $\alpha V_c$ &  \multicolumn{2}{c |}  {}     \\
 \hline
&\multicolumn{2}{c |} {Instability of the first kind
when either} \\
With Coulomb & \multicolumn{2}{c|} {$\alpha$ or $Z\alpha$ is too large}\\
\cline{2-3}
potential $\alpha V_c$ & \multicolumn{2}{c|}{Stability of the second kind
when}\\
& \multicolumn{2}{c| }{both $\alpha$ and $Z\alpha$ are small enough}\\
\hline \hline
\end{tabular}
\bigskip\bigskip

The main point of this paper is the proof of the bottom row of the
second table in the quantized case;  the classical case was done in
\cite{LSS}.

There are several ways in which one could hope to go further.  One is that
one should really prove stability for the {\it binding} energy, i.e.,
one should compute the energy difference between that of free particles
and that of the interacting system.  In a theory with quantized fields
the self-energy, i.e., the energy of a free electron, is unknown and
quite large.  As we show in \cite{selfenergy} and \cite{selfenergy2}
the self-energy of a nonrelativistic particle with spin is bounded below
by $+\Lambda$, and probably even $+\Lambda^{3/2}$.  Moreover, for $N$
fermions (but {\it not} for bosons) this energy is proportional to $C'
N \Lambda$ with $C' >0$.

Another very important problem to consider is renormalization; our mass
$m$ is the unrenormalized one. An answer to this problem also has to
address the question of the meaning of mass in an ultraviolet cut-off
model, since several definitions are possible. Is it the coefficient
of $\beta$ in an effective Dirac operator that gives the renormalized
dynamics, or is it the ground state energy of a ``free'' electron?

The results in this paper will be used in another paper of ours
\cite{binding} to give upper bounds to the hydrogen atom binding energy
(and hence to the mass renormalization using the first definition) in this
relativisitic no-pair model and in some non-relativistic models with
quantized fields.

Finally, let us note that the inclusion of positrons into the model
cannot change the fact that defining an electron by means of $D(0)$
will still cause the instabilities listed in the tables above. The
reason is simply that the existence of positrons does not prevent one
from considering states consisting purely of electrons, and these alone
can cause the listed instabilities.

The use of $D(A)$ instead of $D(0)$ to define the electron requires
a significant change in the Hilbert space structure of QED. It is 
no longer possible to separate the Hilbert space for the electron
coordinates from the Hilbert space (Fock space) of the photons. 
The two are now linked in a manner that we describe in the next section.

\section{Basic Definitions}\label{basic}

We consider $N$ relativistic electrons in the field of $K$ nuclei,
fixed at the positions $R_1,...,R_K\in \R^{3}$. (In  the real world the
fixed nuclei approximation is a good one since the masses of the nuclei
are so large compared to the electron's mass.)  We assume that their
atomic numbers $Z_1,...,Z_K $  are all less than some fixed number $Z
> 0$.  Since the energy is a concave function of each $Z_j$ separately,
it suffices, for finding a lower bound, either to put $Z_j=0$, i.e.,
to remove the $j$-th nucleus, or to put $Z_j=Z$ (see \cite{DL}).  Thus,
without loss of generality, we may assume that all the nuclear charges
are equal to $Z$.

We use units in which $\hbar = 1$ and $c =1$.
$\alpha = e^2/\hbar c$ is the dimensionless ``fine
structure constant'' (=1/137 in nature). The electric charge of the
electron in these units is $e= \sqrt{\alpha}$. 

We use the Coulomb, or radiation gauge so that the Coulomb potential is a
function only of the coordinates of the $N$ electrons, $x_1,x_2,\dots,x_N$ 
and equals
$\alpha V_c$, where
\begin{equation}
V_c = -Z \sum_{i=1}^N \sum_{k=1}^K \frac{1}{|x_i - R_k|} 
+\sum_{1\leq i < j \leq N}\frac{1}{|x_i-x_j|} + 
Z^2\sum_{1\leq k < l \leq K}\frac{1}{|R_k-R_l|} \ .
\end{equation}
In this gauge, it is the vector potential that is quantized. 
A careful discussion of the field and its quantization is given in 
Appendix \ref{appendix:units}. The 
(ultraviolet cutoff) magnetic vector potential
is defined by
\begin{equation}
A(x) = \frac{1}{2\pi} \sum_{\lambda=1}^2 \int_{|k|\leq \Lambda}
\frac{\varepsilon_\lambda(k)}{\sqrt{\omega(k)}} \left(
a_\lambda(k) e^{ik\cdot x} + a_\lambda^{\ast}(k) e^{-ik\cdot x}\right)dk
\  ,
\label{apot}
\end{equation}
where $\Lambda$ is the ultraviolet cutoff on the wave-numbers $|k|$. 
The operators $a_{\lambda}, a^{\ast}_{\lambda}$
satisfy the usual commutation relations
\begin{equation}
[a_{\lambda}(k), a^{\ast}_{\nu} (q)] = \delta ( k-q)
\delta_{\lambda, \nu}\ , ~~~ [a_{\lambda}(k), a_{\nu} (q)] =
0, \quad {\rm{etc}}
\end{equation}
and the vectors $\varepsilon_{\lambda}(k)$ are the two possible
orthonormal polarization vectors perpendicular to $k$ and to each other. 

Our results hold for all finite $\Lambda$. The details of the cutoff in
(\ref{apot})  are quite unimportant, except for the requirement that rotation
symmetry in $k$-space is maintained. E.g., a Gaussian cutoff can be used
instead of our sharp cutoff. We avoid unnecessary generalizations. The
cutoff resides in the $A$-field, not in the field energy,
$H_f$, sometimes called $d\Gamma(\omega)$, which  is given by
\begin{equation}\label{eq:fielden}
H_f = \sum_{\lambda=1,2} ~ \int_{\R^3} ~ \omega(k)
\cre(k) \an(k) d k
\end{equation}

The energy of a photon is  $\omega(k)$ and the physical value of
interest to us, which will be used in the rest of this paper, is
\begin{equation}
\omega(k) = |k|
\end{equation}
Again, generalizations are possible, but we omit them.

An important fact for our construction of the 
physical Hilbert space of our model is that  $[A(x),A(y)]=[B(x),B(y)]=
[A(x),B(y)] =0$
for all $x,y$. Here, $B$ is the magnetic field given by
\begin{equation}
B(x)=\mathrm{curl}\, A(x) =  \frac{i}{2\pi} 
\sum_{\lambda=1}^2 \int_{|k|\leq \Lambda}
\frac{k\wedge \varepsilon_\lambda(k)}{\sqrt{\omega(k)}} \left(
a_\lambda(k) e^{ik\cdot x} - a_\lambda^{\ast}(k) e^{-ik\cdot x}\right)dk
\ . 
\label{bpot}
\end{equation}

The kinetic energy of an electron is defined in terms of a Dirac operator with
the vector potential $A(x)$ (with $x$ being the electron's coordinate) 
\begin{equation} \label{Diracop}
D (A) :={\mbox{\boldmath$\alpha$} }\cdot(- i\nabla +
\sqrt{\alpha} A(x)) +m \beta \ ,
\end{equation}
with ${\mbox{\boldmath$\alpha$}}$  and $\beta$ 
given by the $2\times 2$ Pauli matrices and
$2\times 2$ identity ${\mathrm{I}}$ as

$$
 {{\mbox{\boldmath$\alpha$}}}=
  \left(\begin{array}{cc} 0&\sigma \\
\sigma&0\end{array}\right),\ 
{{\beta}}=
  \left(\begin{array}{cc} {\mathrm{I}}&0 \\
 0&-{\mathrm{I}}\end{array}\right),\ 
\sigma_1=\left(\begin{array}{rr} 0&1 \\ 1&0 \end{array}\right),\ 
\sigma_2=\left(\begin{array}{rr} 0&-i \\ i&0 
\end{array}\right),\
\sigma_3=\left(\begin{array}{rr} 1&0 \\ 0&-1 \end{array}\right) \ .
$$

Note that
\begin{equation}\label{dsquare}
D(A)^2 = \widehat{T}^P(A)(A) + m^2 \ ,
\end{equation}
where $\widehat{T}^P(A)= \left(\begin{array}{cc} T^P(A)&0 \\ 0&T^P(A) 
\end{array}\right)$ 
and $T^P(A)$ is the Pauli operator on $L^2(\R^3; \C^2)$,
\begin{equation}\label{Pauliop}
T^P(A) = \left[ \sigma \cdot (p+\sqrt{\alpha}A(x)) \right]^2
=(p+\sqrt{\alpha}A(x))^2 + \sqrt{\alpha}\, \sigma \cdot B(x)  \ .
\end{equation}    

As a step towards defining a physical Hamiltonian for our system of
$N$ electrons and $K$ fixed nuclei, we first define a conventional, but 
fictitious Hamiltonian 
\begin{equation} \label{hamprime}
H'_N= \sum_{i=1}^N D_i(A) + \alpha V_c  +  H_f  \ .
\end{equation}

This $H'_N$ acts on the usual Hilbert
space $\mathcal{H}_N = \bigotimes^N L^2(\R^3;\C^4)  \bigotimes \mathcal{F}$,
where $\mathcal{F}$ is the Fock space for the $A$-field. A vector in 
$\mathcal{H}_N$ can be written 
\begin{equation}\label{psi}
\Psi = \bigoplus_{j=0}^\infty  \Phi_j(x_1,...,x_N;\,
\tau_1,...,\tau_N;\, 
k_1,...,k_j;\, \lambda_1,...,\lambda_j) \ .
\end{equation}
Here, the $\lambda_i$ take the two values $1,2$ and the the $\tau_i$
take the four values $1,2,3,4$. Each $\Phi_j$ is symmetric in the pairs
of variables $k_i, \lambda_i$ and it
is square integrable in $x,k$. The sum of these 
integrals (summed over $\lambda$'s, $\tau$'s, and $j$)
is finite. The operators $a_\lambda(k)$ and their adjoints act, as
usual, by
\begin{equation}
a_\lambda(k)\Psi =  \bigoplus_{j=0}^\infty  \sqrt{j+1} \Phi_{j+1}(x_1,...,x_N;\,
\tau_1,...,\tau_N;\, 
k_1,...,k_j,k;\, \lambda_1,...,\lambda_j,\lambda) \ .
\end{equation}

As mentioned before, the physical Hilbert space is constructed using the positive spectral
projections of the Dirac operators $ D_i(A)$. By Lemma \ref{unitary}  the
$N$ Dirac operators commute in the strong sense that their spectral projections
commute with each other. Thus, the Hilbert space $\mathcal{H}_N$ can
be divided into $2^N$ subspaces according to the positive and negative spectral
subspaces of each $ D_i(A)$. (Note that as long as $m>0$ there is
no zero spectral subspace.) We denote by $P^+$ the orthogonal projection
onto the positive spectral subspace for {\it all} the Dirac operators.

The space $P^+\mathcal{H}_N$ is invariant (up to unitary equivalence) by
the natural action of the permutation group $S_N$ consisting of permutations
of the electron
labels. 
In accordance with the Pauli principle we choose
the antisymmetric component of $P^+\mathcal{H}_N$, as the physical Hilbert space. 
Thus, our
physical Hilbert space is given as
\begin{equation}\label{physhil}
\mathcal{H}^{\mathrm {phys}}_N= \mathcal{A}\, P^+\mathcal{H}_N
\end{equation}
where $\mathcal{A}$ is the projector onto the antisymmetric component.

Formally, i.e., without attention to domain questions,
our physical Hamiltonian on $\mathcal{H}^{\mathrm {phys}}_N$ is defined to be
\begin{equation}\label{physham}
H^{\mathrm{phys}}_N=P^+ H'_N P^+ \ . 
\end{equation}

Since we are interested in this operator as a quadratic form, it suffices
to specify a domain ${\mathcal{Q}}_N$ which is dense in 
$\mathcal{H}^{\mathrm {phys}}_N$ and on which the expectation values of
the all the operators involved are finite. 
Since all the operators are symmetric, and since a stability estimate entails
that the quadratic form is bounded below, its closure exists and defines a selfadjoint
operator $H^{\mathrm{phys}}_N$. Such a domain ${\mathcal{Q}}_N$ is constructed
in Appendix \ref{projections}. Note that by definition ${\mathcal{Q}}_N$ consists
of antisymmetric elements.

We note that each of the $N$ Dirac operators commute with $P^+$. For $\psi
\in {\mathcal{Q}}_N$  we have  $D_i(A)\psi = P^+ D_i(A)\psi$. For the
other two terms in (\ref{hamprime}) the role of the projector is not so
trivial and that is why we have to write $P^+ H'_N P^+$.
 
This  model has its origins in the work of Brown and Ravenhall
\cite{BrownRavenhall1951} and Sucher \cite{Sucher1980}. The immediate
antecedent is \cite{LSS}.

Let us note five things:

(i).  It is not entirely easy to think about $\mathcal{H}^{\mathrm
{phys}}_N$ because the electronic $L^2$-spaces and the Fock space are now
linked together. In our choice of positive energy states, the electrons
have their own photon cloud. We chose to apply the projector $P^+$ first
and then antisymmetrize.  As explained in Appendix \ref{projections},
we can, of course, do it the other way around and obtain the same Hilbert
space, since $P^+$ commutes with permutations.  We also show in  Appendix
\ref{projections} that $\mathcal{H}^{\mathrm {phys}}_N$ is not trivial;
in fact it is infinite dimensional.

(ii). Usually, in quantum electrodynamics, one defines $\mathcal{H}^{\mathrm {phys}}_N$
by means of the positive spectral subspace of the {\it free} Dirac
operator $D(0)={-i \mbox{\boldmath$\alpha$}} \cdot \nabla + m\beta $,
instead of $D(A)$. This is easier to think about but, as demonstrated in
\cite{LSS} with a classical $A$ field instead of a quantized field, the
choice  of $D(0)$  always leads to instability, as listed in the tables in
Section \ref{intro} and discussed in detail in Appendix \ref{sec:instability}.

(iii). Because of the restriction to the positive spectral subspace of
$D(A)$, the Dirac operator is never negative.  The only negative terms
in $H^{\mathrm{phys}}_N$ come from the Coulomb potential. It should also be
noted that the choice of the free Dirac operator to define an electronic
wave function is not a gauge covariant notion. The $D(A)$ choice {\it is}
gauge covariant.

(iv)  $\mathcal{H}^{\mathrm {phys}}_N $ depends on $\alpha$ and $m$.

(v) While energy, being one component of a four-vector, is not a
relativistically invariant quantity, it is true, nevertheless, that
positive and negative energies of  $D (A)$ are relativistic concepts
since they are invariant under Lorentz transformations that do not change
the direction of time. We thank J-M. Graf for this remark and we thank
J. Yngvason for noting that for this to be true it is essential that
the joint spectrum of energy and momentum of $D (A)$ lies in the light
cone. We have not proved this, but it is plausibly true.

Our main result, to be proved in Section \ref{mainthm}, is
\begin{thm}[Relativistic Quantum electrodynamic Stability]
\label{mainthm}
Assume that $Z$ and $\alpha$ are such that there is a solution $\kappa$ and
$\varepsilon \geq 0$
to the three inequalities (\ref{needs3}), (\ref{needs1}) and (\ref{needs2}).
Then $H^{\mathrm phys}_N$ in (\ref{physham}) 
is bounded below by
\begin{equation}\label{lowerboundmain}
H^{\mathrm phys}_N\geq 
+\sqrt{\varepsilon}\  m \, N - 
 \frac{18\Lambda }{ \pi}  K C_2^3 \ ,
\end{equation}
where 
\begin{equation}
C_2^4 =  \frac{N}{K}\frac{ 6\sqrt{1-\varepsilon} +(\alpha/2)(\sqrt{2Z}+2.3)^2}
{27/2\pi} \ .
\end{equation}
In particular, $Z\leq 42$ is allowed when $\alpha =1/137$. 
\end{thm}

Actually, our proof of Theorem \ref{mainthm} utilizes the absolute value
of the Dirac operator $|D(A)|$ on the Hilbert space  ${\mathcal A}\,
{\mathcal H}_N$. If we recall the connection between $D(A)$ and the
Pauli operator in (\ref{dsquare}) and (\ref{Pauliop}), we can prove 
the following theorem as a byproduct of our proof of
Theorem \ref{mainthm}. Note $\C^2$ in place of $\C^4$ here.

\begin{thm}[Stability with the Pauli operator]  \label{absstab}
Let $H''_N =\sum_{i=1}^N \sqrt{T^P(A) +m^2}\,  + \alpha V_c  +  H_f$  
be a  Hamiltonian on the space  ${\mathcal A}\, \bigotimes^N L^2(\R^3;\C^2)  
\bigotimes \mathcal{F}$. 
Then stability of the second kind
holds under the conditions stated in Theorem \ref{mainthm},
and with the same lower bound (\ref{lowerboundmain}).
\end{thm}

\section{Bounding the Coulomb Potential by a Localized Relativistic Kinetic 
Energy}\label{bounding}

The following Theorem \ref{Coulombbound} contains the main technical
estimate needed in this paper, but it is independently interesting.
It deals with a model of relativistic electrons interacting with quantized
fields, but without the spin-field interaction and without the field
energy. While this model is different from the no-pair Hamiltonian
(\ref{physham}), some of its properties will be useful later.
We consider two such Hamiltonians: A usual one
\begin{equation} \label{relham}
\kappa \sum_{i=1}^N  |p_i+ \, \sqrt{\alpha} A(x_i)| +  V_c  \ ,  
\end{equation}
(with $\kappa >0$) and a related one with a localized kinetic energy
described below in (\ref{kyew}), (\ref{hloc}). In this section $A(x)$ is
some given {\it classical} field, not necessarily divergence free. There
is no $\alpha$ in front of $V$ in (\ref{relham}). 
The Hilbert space is $\mathcal{A}\,
\bigotimes^N\!\! L^2(\R^3;\C^q)$ for fermions with $q$ `spin states'.

With  the  $K$ nuclei positioned at distinct
points $R_j \in \R^3$, for $j=1, \dots , K$, we
define the corresponding Voronoi cells by
\begin{equation}
\Gamma_j = \{x \in \R^3: |x-R_j| <|x-R_i|, i=1, \dots, K, i \not= j \} \ .
\end{equation}
These Voronoi cells are open convex sets.
We choose some $L>0$ and define the balls ${\cal B}_j \subset \R^3 $ by
\begin{equation}
{\cal B}_j = \{x : |x-R_j| \leq 3L \}  \ , 
\end{equation}
and denote by ${\cal B}$ the union of these $K$ balls and by
$\upchi_{{\cal B}} $ the characteristic function of ${\cal B}$.
Similarly, we define smaller balls, 
${\cal S}_j =  \{x : |x-R_j| \leq 2L\}$, and define
$\upchi_{{\cal S}} $ to be the characteristic function of the union of 
these $K$ smaller balls.
Choose some function $g \in W^{1,1}(\R^3)$ with support in
$\{x : |x| \leq 1 \}$, with $g\geq 0$ and with $\int g =1$. 
Define  $g_{L}(x) = L^{-3} g(x/L)$. Clearly $\int g_{L} =1$
and $g_{L} $ has support in  $\{x : |x| \leq L \}$. 
With  $*$ denoting convolution, set
\begin{equation}
\phi_1(x) = g_{L} * \upchi_{{\cal S}} (x) \ .
\end{equation}
This function $\phi_1$ is nonnegative and everywhere bounded by $1$. We
also define $\phi_2 = 1-\phi_1$ and set 
\begin{equation}
F=\phi_1 / \sqrt{\phi_1^2
+ \phi_2^2}, \qquad {\mathrm {and}} \qquad 
G=\phi_2 / \sqrt{\phi_1^2 + \phi_2^2} \ .
\end{equation}
Note that $\phi_1^2 + \phi_2^2 \geq 1/2$ and $F(x)=1$ if $|x-R_j| <
L$ for some $j$. Note also that $F$ and $\phi_1$ are supported in 
${\cal B}$, i.e., $\upchi_{{\cal B}} \phi_1 =\phi_1 $ and
$\upchi_{{\cal B}} F =F$. 
 
We find that 
\begin{equation}  \label{laplacebound}
|\nabla F|^2 + |\nabla G|^2 \leq 4 |\nabla \phi_1|^2 \leq 
\frac{4}{L^2} \left( \int_{\R^3} | \nabla g(x)| dx \right)^2 \ .
\end{equation}
and hence $|\nabla F|, |\nabla G| \leq 2 |\nabla \phi_1|$.

The function $g$ that minimizes the integral in (\ref{laplacebound}) is
$g(x) = 3/4\pi$ for $|x|\leq 1$ and zero otherwise. 
(Although this $g$ is not in $W^{1,1}(\R^3)$ it is a  limit of 
$W^{1,1}(\R^3)$ functions.)
Then the integral equals
$3$  and $|\nabla F|^2 + |\nabla G|^2  \leq 36 L^{-2}$.

The localized kinetic energy  operator $Q(A)$ is given by
\begin{equation}\label{kyew}
Q(A) = F(x)\, |p+\sqrt{\alpha}A(x)|\, F(x) =  
F(x)\sqrt{(p+\sqrt{\alpha}A(x))^2} \, F(x)\ .
\end{equation}
This operator is well defined as a quadratic form since the function $F$
is smooth, and hence defines a self adjoint operator via the Friedrichs
extension.

The related relativistic Hamiltonian, with localized kinetic energy, 
is now defined by
\begin{equation}\label{hloc}
H^{\mathrm{loc}}_N := \kappa \sum_{i=1}^N Q_i(A) + V_c  \ ,
\end{equation}
and has the following bound which, it is to be noted, does not depend
on the details of $g(x)$.

\begin{thm}[Bound on Coulomb energy] \label{Coulombbound}
For any vector field $A(x)$ and for $N$ fermions with $q$ spin states,
\begin{equation}
\kappa \sum_{i=1}^N Q_i(A) + V_c \geq - \frac{N}{2L} 
\max \{(\sqrt {2Z} + 1)^2, \  2Z + \frac{110}{21}  \}  \geq 
-\frac{N}{2L}(\sqrt {2Z} + 2.3)^2 \  ,
\end{equation}
provided  $\kappa \geq \mathrm{max}\{q/0.031,\ \pi Z\}$. 
\end{thm}

\begin{proof}
It was proved in \cite{LY} (eqns. (2.4-2.6) with $\lambda =10/11 $)
that the Coulomb potential $V_c$ is bounded
below by a {\it single-particle} potential plus a constant, namely,
for $x_i,\, R_j \, \in \R^3$, 
\begin{equation}
V_c \geq - \sum_{i=1}^N W(x_i) + \frac{Z^2}{8} \sum_{j=1}^K \frac{1}{D_j}
\label{coulomblower}
\end{equation}
where $2D_j= \min_{i\neq j}\{|x_i-x_j|\}$ and, for $x \in \Gamma_j$,
\begin{eqnarray}
W(x) & = & \frac{(\sqrt Z + 1/\sqrt 2\ )^2}{|x-R_j|} \ \  \mathrm{for}\ 
|x-R_j|\geq  \frac{10 D_j}{11}   \nonumber \\
 & = & \frac{Z}{|x-R_j|}+\frac{121}{42 D_j} \ \  \mathrm{for}\  |x-R_j|<
\frac{10 D_j}{11} \ .
\end{eqnarray}
This estimate reduces our problem to finding a lower bound to
\begin{equation}\label{equivcoul}
\kappa \sum_{i=1}^N Q_i(A)- \sum_{i=1}^N F(x_i)^2W(x_i) 
- \sum_{i=1}^N(1- F(x_i)^2)W(x_i) + 
\frac{Z^2}{8} \sum_{j=1}^K \frac{1}{D_j} \ .
\end{equation}
Since $F(x)=1$ if $|x-R_j| <L$ for some $j$, the third term in
(\ref{equivcoul}) is bounded below by
\begin{equation}
-\frac{N}{2L} \max \{(\sqrt {2Z} + 1)^2, \ 2Z + \frac{110}{21} \} \  .
\end{equation}
Estimating the first and second terms using the Pauli exclusion principle
amounts to filling  the lowest possible energy levels with $q$ electrons
each, and this energy is bounded below by $q$ times the sum of the
negative eigenvalues of the operator
\begin{equation}
F(x)(|p+\sqrt{\alpha}A(x)| - W(x))F(x) \ .
\end{equation}

According to the generalized min-max principle \cite{Anal} Corollary
12.2, and the fact that $\Vert F\psi \Vert \leq \Vert \psi \Vert$, this
is bounded below by  $q$ times the sum of the negative eigenvalues
of the operator $|(p+\sqrt{\alpha}A(x))| -W(x)$.  However, 
Theorem 1 of \cite{LY},
shows that this sum is not less than $(-Z^2/8) \sum_{j=1}^K
1/D_j$ under the stated condition on $\kappa$.

(Notes:   We refer here to Theorem 1 of \cite{LY} because, as noted
in \cite{LSS}, the proof of that theorem holds for $|p+\sqrt{\alpha}A(x)|$ 
in place  of $|p|$.  
 While Theorem 1 of \cite{LY} is stated in terms of $V_c$,
the proof in \cite{LY} actually replaces $V_c$ by its lower bound
(\ref{coulomblower}). ) \end{proof}

\section{Proof of Theorem \ref{mainthm}}   \label{nopair}

We employ a strategy similar to that in \cite{LSS}. 

As a first step we use Theorem \ref{Coulombbound} with a suitable choice of 
$L$ to control the
Coulomb potential.

The operators appearing in Theorem \ref{Coulombbound} do not involve spin,
but the number of spin states, $q$,  is important for determining the
relevant value of $\kappa$. The correct choice is $q=2$, not $q=4$, as
explained in \cite{LSS} page 42 and appendix B. The point is the following. The one-body
density matrix $\Gamma (x, \sigma ; x', \sigma ')$ coming from an 
antisymmetric $N$ particle wave function $\Psi$ defines
a reduced one body density matrix 
\begin{equation}
\gamma(x,x') = \sum_{\sigma = 1}^4 \Gamma (x, \sigma ; x', \sigma ) \ .
\end{equation}
This reduced density matrix, in general, satisfies $0 \leq {\rm Tr}\gamma 
\equiv \int\gamma(x,x) dx \leq 4$.
If, however, $\Psi$ is in the range of $P^+$, then $0 \leq {\rm Tr} \gamma \leq 
2$,
as shown in \cite{LSS}. In the proof of Theorem \ref{Coulombbound}, the only
relevant information about $\Psi$ enters via the reduced single particle
matrix $\gamma$. Thus, we require only 
$\kappa \geq \mathrm{max} \{64.5, \pi Z \}$.

In the definition of $F$  we set $L= C_2/\Lambda$ where
$C_2>0$ is some constant to be conveniently chosen later. We then have
(recalling (\ref{dsquare}), (\ref{Pauliop}), and $P_i^+ D_i(A)P_i^+ =
P_i^+|D_i(A)|P_i^+$)
\begin{equation} \label{relpauli}
P^+\left[ \sum_{i=1}^N D_i(A)+ \alpha V_c \right]P^+\geq 
P^+ \sum_{i=1}^N \left[\sqrt{\widehat{T}_i^P(A) + m^2}-\kappa \alpha  Q_i(A) 
\right]P^+  - 
\alpha N\frac { \Lambda}{2C_2} (\sqrt {2Z} + 2.3)^2  P^+\  .
\end{equation}
(Here, $Q(A)$ really denotes the $4\times 4$ operator $Q(A)\otimes I_4$ where 
$I_4$ is the identity in spin-space.)
Consider the operator
\begin{equation}\label{h2}
H_2:= P^+ \sum_{i=1}^N \left[\sqrt{\widehat{T}_i^P(A) + m^2}-\delta m -
\kappa \alpha Q_i(A)
+{C_3} \Lambda \right]P^+ \ ,
\end{equation}
where the numbers $0\leq \delta  \leq 1$ and $C_3 >0 $ will be chosen  later. 

If we denote by $\Xi^+$ the projection onto the positive spectral subspace of
$D(A)$ acting on $L^2(\R^3;\C^4)\bigotimes {\mathcal F}$, 
then $H_2$ is bounded below by
\begin{equation}
{\rm Tr}_4[\Xi^+ S\Xi^+]_-   \ ,
\end{equation}
where ${\rm Tr}_n$ with $n=1,2,4$ denotes the trace on $L^2(\R^3;\C^n)$.
The operator $S$ is
\begin{equation}
S:=\sqrt{\widehat{T}^P(A) + m^2}- \delta m - \kappa \alpha  Q_i(A) +
{C_3}\Lambda \  .
\end{equation} 
It has the form
\begin{equation}
S= \begin{pmatrix} Y& 0 \\ 0 & Y
\end{pmatrix} \ .
\end{equation}
Here, the entry $Y$ is a $2 \times 2$ matrix valued operator and $[X]_-$
denotes the negative part of a self-adjoint operator $X$ (and which is
nonnegative by definition).  The projection $\Xi^+$ is not explicitly
given, but observing, as in \cite{LSS}, that the projection $\Xi^-$
onto the {\it negative} energy states is related to $\Xi^+$ by 
\begin{equation}
\Xi^- = U^{-1}\Xi^+U = -U\Xi^+U \ ,
\end{equation}
where $U$ is the matrix
\begin{equation}\label{you}
U=
\begin{pmatrix} 0 & {\mathrm{I}} \\ -{\mathrm{I}} & 0 
\end{pmatrix} \ ,
\end{equation}
we see that the operators $\Xi^+S\Xi^+$ and $\Xi^-S\Xi^-$
have the same spectrum.
Thus, 
\begin{equation}
{\rm Tr}_4[\Xi^+S\Xi^+]_-  \leq  \frac{1}{2} {\rm Tr}_4[S]_- = 
{\rm Tr}_2\left[ \sqrt{T^P(A) + m^2}-\delta m -\kappa \alpha Q(A)
+C_3\Lambda \right]_-  \ .
\end{equation}
Therefore, the infimum of the spectrum of $H_2$ over states that satisfy the 
Pauli exclusion principle (with $4$ spin states) is bounded below by
\begin{equation}
- {\rm Tr}_2\left[ \sqrt{T^P(A) + m^2}-\delta m -\kappa \alpha Q(A) + {C_3}\Lambda
\right]_-   \ . 
\end{equation}

The BKS inequality \cite{BKS} (see also \cite{LSS}) states that
for positive operators $A$ and $B$, 
${\rm Tr}_2[A-B]_- \leq {\rm Tr}_2[A^2-B^2]_-^{1/2}$. Note that 
$\sqrt{T^P(A) + m^2}-\delta  m \geq 0$ and,  therefore,
\begin{equation}
H_2 \geq -{\rm Tr}_2\left[\left(\sqrt{T^P(A) + m^2}-\delta  m +{C_3}
\Lambda \right)^2  -\kappa^2 \alpha^2 Q(A)^2 \right]_-^{1/2}
\end{equation}
which is greater than
\begin{equation}\label{great}
-{\rm Tr}_2\left[\left( \sqrt{T^P(A)
 +m^2} -\delta  m \right)^2 + C_3^2\Lambda^2  
-\kappa^2 \alpha^2 Q(A)^2 \right]_-^{1/2} \ .
\end{equation}
(Here, and in the following, we use the fact that ${\rm Tr} [X]_-^{1/2}$ is
monotone decreasing in $X$.)

Next, we expand $( \cdots )^2$ in (\ref{great}) and use the arithmetic-geometric 
mean inequality to bound (\ref{great}) from below by
\begin{equation} \label{small}
-{\rm Tr}_2\left[(T^P(A) +m^2)(1-\varepsilon)+ (1-1/\varepsilon)m^2 \delta^2
+ C_3^2\Lambda^2   -\kappa^2 \alpha^2 Q(A)^2 \right]_-^{1/2} \ .
\end{equation}

We choose $\delta$ so that the mass disappears, i.e., $\delta^2=\varepsilon$.

The next step is to  localize the Pauli term $T^P(A)$. 
A standard calculation shows that (with $F,G$ as in Section \ref{bounding})     
\begin{equation}\label{tee}
T^P(A)= FT^P(A)F + GT^P(A)G -|\nabla F|^2 -|\nabla G|^2
\geq  FT^P(A)F -|\nabla F|^2 -|\nabla G|^2 \  .
\end{equation}
We insert the right side of (\ref{tee}) into (\ref{small}) and, 
recalling (\ref{laplacebound}), choose $C_3$ to eliminate the
$\Lambda^2$ term, i.e.,
\begin{equation}\label{see3}
C_3 = \frac{2 \sqrt{1-\varepsilon}}{ C_2} \left(\int_{\R^3} 
|\nabla g(x)|dx \right) =   \frac{6\sqrt{1-\varepsilon}}{ C_2} \ .
\end{equation} 

Thus, using the fact that $ Q(A)^2 = F |p+\sqrt{\alpha}A(x)|F^2 |p+\sqrt{\alpha}A(x)| 
F \leq F (p+\sqrt{\alpha}A(x))^2 
F$, which follows from
$F^2\leq 1$, we obtain the bound
\begin{align}
H_2 &\geq -{\rm Tr}_2\left[(1-\varepsilon)FT^P(A)F  -\kappa^2\alpha^2 Q(A)^2 
\right]_-^{1/2} \notag \\
&\geq -{\rm Tr}_2\left[F\left((1-\varepsilon -\kappa^2\alpha^2) (p+\sqrt{\alpha}A(x))^2 + 
 (1-\varepsilon)\sqrt{\alpha}\upchi_{{\cal B}}\sigma \cdot B(x) \right) F 
\right]_-^{1/2} \ . \label{paulibound}
\end{align}
We have used the fact that $\upchi_{\cal B}F =F$.

Since $F X F \geq - F[X]_-F$ for any $X$,  the eigenvalues 
of $FXF $ are bounded below by the eigenvalues of $-F[X]_-F$, and
hence we have that 
${\rm Tr}\left[F XF  \right]_-^{1/2} \leq 
{\rm Tr}\left[F X_- F \right]^{1/2}$, and hence
\begin{multline}
{\rm Tr}_2 \left[F\left\{(1 - \varepsilon -\kappa^2\alpha^2)(p+\sqrt{\alpha}A(x))^2 + 
(1-\varepsilon)\sqrt{\alpha}\upchi_{{\cal B}}\sigma 
\cdot B(x)\right\}F \right]_-^{1/2}  \nonumber \\
\leq  {\rm Tr}_2\, \left\{F\left[(1-\varepsilon -\kappa^2\alpha^2)(p+\sqrt{\alpha}A(x))^2 
+ 
(1-\varepsilon)\sqrt{\alpha}\upchi_{{\cal B}}\sigma 
\cdot B(x)\right]_-F\right\}^{1/2} \ .
\end{multline}
The expression $[\quad ]_-$ between the two $F$'s is, by definition, 
a positive-semidefinite self-adjoint operator and we denote it
by $Y$. Now
\begin{equation}
{\rm Tr}_2 (FYF)^{1/2} = {\rm Tr}_2 (FY^{1/2} Y^{1/2}F)^{1/2} = 
{\rm Tr}_2 (Y^{1/2}F FY^{1/2})^{1/2}
\end{equation}
since, quite generally, $X^*X$ and $XX^*$ have the same spectrum (up to
zero eigenvalues, which are not counted here).
Finally, we note that since $F^2 \leq 1$,  $Y^{1/2}F FY^{1/2} \leq Y$,
and hence
${\rm Tr}_2 (FYF)^{1/2}= {\rm Tr}_2 (Y^{1/2}F FY^{1/2})^{1/2} 
\leq {\rm Tr}_2 Y^{1/2}$.
Thus, it remains to find an upper bound to $[h]_-^{1/2}$ where
\begin{equation}
h= (1-\varepsilon -\kappa^2\alpha^2)(p+\sqrt{\alpha}A(x))^2 + 
(1-\varepsilon)\sqrt{\alpha}\upchi_{{\cal B}}\sigma \cdot  B(x)  \   .
\end{equation}

Denote the negative eigenvalues of $h$
by $-e_1 \leq -e_2 \leq \cdots $.
One way to bound the eigenvalues from below  is to replace $\sigma\cdot
B(x)$ by $-|B(x)|$, but then each eigenvalue of $\widetilde{h} := 
(1-\varepsilon -\kappa^2\alpha^2)(p+\sqrt{\alpha}A(x))^2 
-(1-\varepsilon)\sqrt{\alpha}\upchi_{{\cal B}} |B(x)|$ on $L^2(\R^3)$ would  have to be 
counted twice (because ${\mathrm {Tr}}_2$ is over $L^2(\R^3;\C^2)$ and not
$L^2(\R^3)$). As shown in \cite{Loss}, however, the intuition that each 
negative eigenvalue of $\widetilde{h}$ should be counted only once is correct.
Thus, ${\mathrm {Tr}}_2 [h]_-^{1/2} \leq {\mathrm {Tr}}_1 [\widetilde{h}]_-^{1/2}$. 
By the Lieb-Thirring inequality \cite{LiebThirring1976}
we obtain the bound
\begin{equation} \label{lt}
\sum_i \sqrt{e_i} \leq \frac{(1-\varepsilon)^2\ \ell\ \alpha}{(1-\varepsilon -
\kappa^2\alpha^2)^{3/2}} 
\int_{{\cal B}}  B(x)^2 {\rm d}x \ ,
\end{equation}
with $\ell = 0.060$ \cite{Loss}.

It is to be emphasized that (\ref{lt}) is an {\it operator} inequality. 
That is, the operator in (\ref{h2}), which is part of $H_N^{\mathrm
{phys}}$, satisfies
\begin{equation} \label{hbound}
H_2 \geq - \frac{(1-\varepsilon)^2\ \ell\ \alpha}{(1-\varepsilon -
\kappa^2\alpha^2)^{3/2}}
\int_{{\cal B}}  B(x)^2 {\rm d}x \ .
\end{equation}

The right side of (\ref{hbound}) can be controlled by the field energy
through inequality (\ref{j.10}) --- provided $1/8\pi $ is not less than 
 the constant in (\ref{lt}), (\ref{hbound}).

\subsection{Evaluation of Constants} We are now ready to list the
conditions on the constants $C_2$ and $ \varepsilon$ that have
been introduced and to use these to verify the results of Theorem
\ref{mainthm}.

\begin{eqnarray} 
 \kappa &\geq& {\mathrm {max}} \{64.5, \, \pi Z \} \label{needs3}\\
{\mathrm{Conditions:}}\qquad \qquad\qquad \qquad
(\kappa \alpha)^2 &<&
1-\varepsilon \leq 1 \label{needs1}\\
\frac{(1-\varepsilon)^2 \alpha}{(1-\varepsilon -
  \kappa^2\alpha^2)^{3/2}} &\leq& \frac{1}{8\pi (0.060)}  \label{needs2}
\end{eqnarray}
The first comes from  Theorem \ref{Coulombbound} with $q=2$.
The second is the condition that the  kinetic energy term in
$H_2$ is positive. The third is the requirement that the the field energy
$H_f$ dominates the sum of the negative eigenvalues in (\ref {lt}).

Assuming these conditions are satisfied the total energy is then bounded below 
by the sum of the following four terms (recalling (\ref{see3})
and $\delta^2=\varepsilon$):
\begin{eqnarray}\label{energies1}
&&+\sqrt{\varepsilon}\  m  \, N\\
&&-\frac{6 \sqrt{1-\varepsilon} } {C_2} \, \Lambda \, N \label{energies2}\\
{\mathrm{ Energy\ Lower\  Bounds:}} &&-\frac{\alpha \, \Lambda}{2C_2} 
\left(\sqrt{2Z} +2.3 \right)^2 \, N \label{energies3}\\
&&-\frac{\Lambda^4}{8\pi^2}\int_{\mathcal B} 1 \geq - \frac{4\pi}{3} 
\frac{\Lambda}{8\pi^2} \left( 3C_2\right)^3 \, K=- \frac{9}{2\pi}\Lambda 
C_2^3 \, K \label{energies4}
\end{eqnarray}
The first comes from the $-\delta m $ term in (\ref{h2}). Similarly,
the second comes from the $+ C_3\Lambda $ term in (\ref{h2}). The third
term is the last term in (\ref{relpauli}) which, in turn, comes from
Theorem \ref{Coulombbound}.  The fourth term is the additive constant in
(\ref{j.10}) with $w(y)= \upchi_{\cal B}(y)$. 
The volume of ${\cal B}$ is bounded by the number of nuclei times
the volume of one ball of radius $3L$ around each nucleus.

Obviously we choose
\begin{equation} \label{see2}
C_2^4 =  \frac{N}{K}\frac{ 6\sqrt{1-\varepsilon} +(\alpha/2)(\sqrt{2Z}+2.3)^2}
{27/2\pi} \ .
\end{equation}
The sum of the terms  (\ref{energies1} -- \ref{energies4}) 
then become our lower bound for the energy
\begin{equation} \label{finalenergy}
\frac{E}{N} \geq  +\sqrt{\varepsilon}\  m  - 
 \frac{18\Lambda }{ \pi}  \frac{K}{N} C_2^3 \  ,
\end{equation}
which  satisfies stability of the second kind. 

To find the largest possible $Z$ for which stability holds we take
$\alpha = 1/137$ and make the choice $\varepsilon=0$. 
We then find,
from (\ref{needs2}), that  $ \kappa \alpha \leq 0.97$. Setting
$\kappa = \pi Z$  we find stability up to
$Z=42$.

The choice $\varepsilon=0$ makes the energy in (\ref{finalenergy}) negative.
Recall that if $Z=0$  then  $E/N= m$. To make  contact with physics we would
like the energy to be positive, i.e., only a little less than $Nm$. 
To fix ideas, let us consider the case $\pi Z \leq 64.5$ and $\alpha = 1/137$.
Then $\kappa =64.5$, $\kappa \alpha =0.471$ and $(\kappa \alpha)^2 =0.222$. 
{}From (\ref{needs2}), we require that (with $x=1-\varepsilon \geq 0.222$)
\begin{equation}
x^2 \leq 90.9 (x-0.222)^{3/2} \ ,
\end{equation}
which means that  we can take $1-\varepsilon = .229$  or $\varepsilon =
0.771$. 

Now let us consider the case of hydrogen, $Z=1$ and $N=K$ (neutrality). 
{}From (\ref{see2})  we find that $C_2 =  0.908$. Then (\ref{finalenergy})
becomes
\begin{equation}
\frac{E}{N}= 0.866 m - 4.29\Lambda  \ . 
\end{equation}
If $\Lambda$ is less than one fifth of the electrons's self-energy, 
the total energy of arbitrarily many hydrogen atoms is positive. This
bound could be significantly improved by more careful attention to
our various inequalities.

\appendix
\section{Appendix: A Note About Units}\label{appendix:units}

The choice of units in electrodynamics is always confusing, especially
when interactions with charged particles are involved. 

The interaction of the magnetic vector
potential with a charged particle is $eA(x)$.
In cgs units the classical field energy is
\begin{equation}\label{bsquare}
H_f^{\mathrm{classical}} = \frac{1}{8\pi} \int_{\R^3}\left\{ B(x)^2
+ E(x)^2\right\} dx \ .
\end{equation}
With $B(x)=\mathrm{curl}A(x)$,
we use the Coulomb (or radiation gauge) so that 
$\mathrm{div}A(x)=0 $ and $\mathrm{div}E(x)=0 $.

We define $a_\lambda(k)$ and its complex-conjugate (classically) or adjoint
(quantum-mechanically), $a_\lambda^{\ast}(k)$, in terms of the Fourier
transform of (the real fields) $A(x)$ and $E(x)$ as follows.
\begin{eqnarray} 
A(x) &=& \frac{\sqrt{\hbar c}}{2\pi} \sum_{\lambda=1}^2 \int_{\R^3}
\frac{\varepsilon_\lambda(k)}{\sqrt{|k|}} \left(
a_\lambda(k) e^{ik\cdot x} + a_\lambda^{\ast}(k) e^{-ik\cdot x}\right)dk
\label{a}\\
E(x) &=&  \frac{i\sqrt{\hbar c}}{2\pi} \sum_{\lambda=1}^2 \int_{\R^3}\sqrt{|k|}
\, \varepsilon_\lambda(k) \left(
a_\lambda(k) e^{ik\cdot x} - a_\lambda^{\ast}(k) e^{-ik\cdot x}\right) dk \ ,
\label{e}
\end{eqnarray}
in terms of which
\begin{equation}
B(x) = \frac{i\sqrt{\hbar c}}{2\pi} \sum_{\lambda=1}^2 \int_{\R^3}
\frac{k\wedge \varepsilon_\lambda(k)}{\sqrt{|k|}} \left(
a_\lambda(k) e^{ik\cdot x} - a_\lambda^{\ast}(k) e^{-ik\cdot x}\right)dk
 \label{b} \ .
\end{equation}
The parameter $\sqrt{\hbar c}/2\pi$ in (\ref{a}--\ref{b}) were chosen
purely for convenience later on.
The two unit vectors here, $\varepsilon_\lambda(k)$, $\lambda = 1,2$,
are perpendicular to each other and to $k$ (which guarantees that $\mathrm{div}
A=0$). They cannot be defined on the whole of $\R^3$ as
smooth functions of $k$ (although they can be so defined with the use of 
`charts'), but that will be of no concern to us.

Thus, when (\ref{a} -- \ref{b}) are substituted in (\ref{bsquare}) 
we obtain
(using Parseval's theorem and \\
$\int \exp({ik\cdot
x})dx=(2\pi)^3 \delta(k)$ and $|k|^2\varepsilon_\lambda(k)\cdot 
\varepsilon_\lambda(-k) = -\left(k\wedge \varepsilon_\lambda(k)\right)     
\cdot \left(-k\wedge \varepsilon_\lambda(-k)\right)$)    
\begin{equation} \label{classenergy2}
H_f^{\mathrm{classical}} =  \frac1{2}\, \hbar c \sum_{\lambda=1}^2 \int_{\R^3}
 |k| \left\{a_\lambda^{\ast}(k)a_\lambda(k)
+ a_\lambda(k)a_\lambda^{\ast}(k) \right\}dk
\end{equation}
(Although $a_\lambda^{\ast}(k)a_\lambda(k) =a_\lambda(k)a_\lambda^{\ast}(k)$
for functions, this will not be so when $a_\lambda(k)$ is an operator.
The form in (\ref{classenergy2}) is that obtained after the substitution
just mentioned.)

To complete the picture, we quantize the fields by making the 
$a_\lambda(k)$ into operators with the following commutation relations.
\begin{equation}\label{commutator}
\big[a_\lambda(k), \,  a_{\lambda^\prime}^{\ast}(k^\prime)\big] =
\delta_{\lambda, \, \lambda^\prime}\delta(k-k^\prime)
\quad \quad \mathrm{and} \quad \quad
\big[a_\lambda(k), \,  a_{\lambda^\prime}(k^\prime)\big]=0 \ .
\end{equation}
The quantized field energy is obtained from (\ref{classenergy2},
\ref{commutator}) and is given by the Hamiltonian operator
\begin{equation}\label{fieldham}
H_f =  \hbar c \sum_{\lambda=1}^2 \int_{\R^3} |k|
a_\lambda^{\ast}(k)a_\lambda(k) dk \  .
\end{equation}
It agrees with (\ref{classenergy2}) up to an additive `infinite
constant'.

In the rest of this paper we omit $\hbar c $ since we use units in
which $\hbar=c=1$. 

\section{Appendix: Field Energy Bound}  \label{sec:appendix}

In this appendix we prove (\ref{j.10}) which relates the localized
classical field energy to the quantized field energy. A proof was given in
\cite{BFG}. The small generalization given here is a slightly modified
version of that in \cite{ selfenergy, selfenergy2}.

Consider a collection of operators (field modes), 
parametrized by $y \in {\R}^3$,  and by $j$ in some set of integers
($j \in \{1,2,3\}$ in our case of interest)
given, formally, by
\begin{equation}
L_j(y) = \sum_{\lambda =1}^2 \int_{|k| < \Lambda} \sqrt{|k|}\,
\widehat{{v}}_{\lambda,j} (k)e^{i k \cdot y}a_{\lambda}(k) dk \ , \label{ell}
\end{equation}
where $\widehat{v}_{\lambda,j} $ is the Fourier transform of 
some arbitrary complex function ${v}_{\lambda,j}(x)$.
Our convention for the Fourier transform of a general function $g(x)$ is
\begin{equation} \label{ftrans}
{\widehat{g}} (k) = (2\pi)^{-3/2}\int_{\R^3} g(x)e^{+i k \cdot x} dx 
\qquad \mathrm{and}\qquad
{g}(x) =  (2\pi)^{-3/2} \int_{\R^3} 
{\widehat{g}} (k)e^{-i k \cdot x} dk \ .
\end{equation}

The following lemma is elementary. It involves 
${v}_{\lambda,j}(x)$  and  a summable function 
$w(x)$, with a  norm defined by
\begin{equation} \label{norm}
\Vert w \Vert_v := \sup_{f_\lambda(x)}\frac{
\sum_j \int_{\R^3} |\sum_{ \lambda}{f_\lambda \ast v_{\lambda,j}(x)}|^2 
\, |w(x)| dx }
{\sum_{ \lambda} \int_{\R^3} |{f_\lambda}(x)|^2 dx }   \  , 
\end{equation}
where $\ast$ is convolution. 
\bigskip

\begin{lm}[Lower bound on field energy] \label{lowerbound}
Assume that 
$\Vert w \Vert_v \leq 1$. Then
\begin{equation} \label{j.3}
H_f \geq \sum_j \int_{\R^3} {w(y) } L^{*}_j(y) L_j(y)  d y \ .
\end{equation}
Moreover, if $w(y)\geq 0$, for all $y$ then
\begin{equation} \label{j.4}
H_f \geq  \frac1{4}
 \sum_j \pm  \int_{\R^3} {w(y) } (L_j(y) \pm L_j^{*}(y))^2  d y 
-\frac1{2}\sum_j \sum_{\lambda =1}^2\int_{|k| < \Lambda}  |k| \,
|\widehat{v}_{\lambda,j} (k)|^2 dk \ \int_{\R^3} w(y) 
 d y  \  ,  
\end{equation}
for any choice of + or - for each $j$.  (Note that $-(L-L^*)^2 \geq 0$.)

\begin{proof} The difference of the two sides in (\ref{j.3})
is a quadratic form of the type  \\
$\sum_{\lambda,\lambda'} \int \int a^*_\lambda (k) Q(k,\lambda :  k',\lambda')
a^{\phantom{*}}_{\lambda'}(k') dk dk'$.  In order to establish (\ref{j.3}) 
it is necessary and sufficient
to prove that the matrix $ Q(k,\lambda :  k',\lambda') $ is positive
semidefinite.  This is the condition that
\begin{equation}
 \sum_{\lambda=1}^2 \int_{|k| < \Lambda} |\widehat{f}_{\lambda}(k)|^2 dk
 \geq  (2\pi)^{3/2} \sum_{j}\sum_{\lambda, \lambda^{\prime}=1}^2 
\int \int_{|k|,|k^{'}| < 
\Lambda}\overline{\widehat{f}_{\lambda}(k){\widehat{v}}_{\lambda,j}(k)}
\widehat{f}_{\lambda^{\prime}} (k^{\prime})
\widehat{{v}}_{\lambda^\prime,j}(k^\prime)
\widehat{w}(k^\prime-k)dk dk^\prime   \label{j.6}
\end{equation}
for all $L^2$ functions $\widehat{f}_{\lambda}(k)$. 
Condition (\ref{j.6}) is just the condition that  $\Vert w \Vert_v \leq 1$,
since $\widehat{f*v}(k) = (2\pi)^{3/2} \widehat{f}(k) \widehat{v}(k)$. 

To obtain (\ref{j.4}) from (\ref{j.3}) we use the three facts that $
w(x) \geq 0$,   that
\begin{equation}
L_j^{*}(y) L_j(y) = L_j(y) L_j^{*}(y)-  \sum_{\lambda =1}^2 \int_{|k| < 
\Lambda}|k| |\widehat{v}_{\lambda,j} (k)|^2 dk\ , \label{j.7}
\end{equation}
and that, quite generally for operators, 
\begin{equation}
\pm \left( LL + L^{\ast} L^{\ast}\right) \leq  L^{\ast}L + L L^{\ast} \ . 
\label{j.8}
\end{equation}
\end{proof}
\end{lm}

\bigskip

The following examples are important. 
First, we define the ultraviolet cutoff fields 
$A^\Lambda, \, B^\Lambda, \,
E^\Lambda$ as in (\ref{a},\ref{b},\ref{e}) except that the $k$
integration is over $|k|\leq \Lambda$ instead of $\R^3$. E.g.,
\begin{equation}
B^\Lambda(x) = \frac{i}{2\pi} \sum_{\lambda=1}^2 
\int_{|k|\leq \Lambda}
\frac{k\wedge \varepsilon_\lambda(k)}{\sqrt{|k|}} \left(
a_\lambda(k) e^{ik\cdot x} - a_\lambda^{\ast}(k) e^{-ik\cdot x}\right)dk
 \label{blambda} \  ,
\end{equation}
recalling that $\hbar =c=1$.

This notation, $A^\Lambda, \, B^\Lambda, \,E^\Lambda$, with the 
superscript $\Lambda$, will be used in this appendix {\it only}.

For the first two examples we define $\widehat{v}_{\lambda,j}(k) =  (
\left( k\wedge \varepsilon_\lambda(k) \right)_j /(2\pi)^{3/2} |k|$, so
that $i(L_j(x)-L_j^{\ast}(x))= B^\Lambda_j(x)/\sqrt{2\pi}$ for $j=1,2,3$.

\medskip

\noindent
{\it Example 1:} Assume  that
$ 0\leq w(y) \leq 1$ for all $y$. Then Lemma \ref{lowerbound} implies 
\begin{equation} \boxed{\quad
H_f \geq  \frac{1}{8\pi}\int_{\R^3}B^{\Lambda}(x)^2 w(x)dx - 
\frac{\Lambda^4}{ 8\pi^2} \int_{\R^3} w(y)\, dy \  . \quad} \label{j.10} 
\end{equation}

To verify the norm condition (\ref{j.6}) we note first
that in view of (\ref{norm}) it suffices to assume
that $w(y)\equiv 1$. Then, $\widehat{w}(k) = (2\pi)^{3/2}
\delta(k)$. On the right side of (\ref{j.6}) we may use the equality
$\sum_j\overline{\widehat{v}_{\lambda,j}(k)}\widehat{v}_{\lambda^\prime,j}(k)
=(2\pi)^{-3} \delta_{\lambda, \, \lambda^\prime}$ (because $( \left(
k\wedge \varepsilon_\lambda(k)/|k|\right)_j$ are the three components
of two orthonormal  vectors). Thus, (\ref{j.6}) is not only satisfied,
it is also an identity with this choice of $\widehat{w}$. Finally,
(\ref{j.4}) is exactly (\ref{j.10}) since $\int_{|k|\leq \Lambda} |k|dk =
\pi \Lambda^4$.

\bigskip

{\it Example 2:} Take $w(y)= C \delta(x-y)$, 
with $x$ some fixed point
in $\R^3$ and where $C>0$ is some constant. Then $\widehat{w}(k) =
C(2\pi)^{-3/2}\exp(ik\cdot x)$ and $|\widehat{w}(k)|=C(2\pi)^{-3/2}$.
We take $\widehat{v}_{\lambda, \, j}(k)$ as in Example 1.
The  right side of   condition (\ref{j.6})
equals $C\sum_j |\int_{|k|\leq \Lambda}\sum_\lambda \widehat{f}_\lambda(k) 
\widehat{v}_{\lambda, \, j}(k)\exp(ik\cdot x)dk|^2$,
and a simple variational argument then says that we should choose
$\widehat{f}_\lambda(k)  = \exp(-ik\cdot x) 
\overline{\widehat{v}_{\lambda}(k)}\cdot V$,
where $V$ is some fixed vector.
By spherical symmetry we can take $\widehat{f}_\lambda(k)=
\overline{\widehat{v}_{\lambda, \, 1}(k)}$,
and  (\ref{j.6}) becomes 
\begin{equation}\label{norm2}
1\geq C\int_{|k|\leq \Lambda}\sum_\lambda |\widehat{v}_{\lambda, \, 1}(k)|^2 dk
=  C \, \frac{2}{3}\cdot  \frac{1}{(2\pi)^3}\cdot  
\frac{4\pi}{3} \Lambda^3 \ .
\end{equation}

With $C= {9}{\pi^2}\Lambda^{-3} $, and with 
$$\sum_\lambda 
\sum_j \int_{|k| < \Lambda}  |k| \ 
|\widehat{v}_{\lambda,j} (k)|^2 dk \  \int_{\R^3} w(y)dy 
= 2\cdot (2\pi)^{-3} \cdot (4\pi/4)
\Lambda^4  C  \  ,
$$
(\ref{j.4}) becomes
$H_f \geq \frac1{4}\cdot C\cdot (2\pi)^{-1}B^\Lambda(x)^2  -
      (1/8)\pi^{-2}\cdot \Lambda^4  \cdot C  $, or 
\begin{equation}\label{j.11}
\boxed{\quad
H_f  \geq  \frac{9\pi}{8}\Lambda^{-3}B^\Lambda(x)^2 -\frac{9}{ 8}\Lambda 
\quad}
\end{equation}

This can also be used \cite{selfenergy, selfenergy2} with $x$ being the electron
coordinate (which is an operator, to be sure, but is one that commutes
with the field operators).

\bigskip

{\it Example 3:}  If we replace $\left( k\wedge \varepsilon_\lambda(k) 
\right)_j/|k|$
in $v_{\lambda,\, j}(k)$ by $(\varepsilon_{\lambda}(k))_j$ then everything
goes through as before
and we  obtain (\ref{j.10}) and (\ref{j.11}) with $E^\Lambda(x)^2$
in place of $B^\Lambda(x)^2 $. 

\bigskip

{\it Example 4:}  We now take $\widehat{v}_{\lambda,j}(k) = 
(\varepsilon_\lambda(k))_j (2\pi)^{-3/2} /|k|$, so
that $(L_j(x)+L_j^{\ast}(x))= A^\Lambda_j(x)/\sqrt{2\pi}$ for $j=1,2,3$.
The analysis proceeds as in Example 2, except that the normalization
condition (\ref{norm2}) becomes
$$
1\geq C\int_{|k|\leq \Lambda}\sum_\lambda |v_{\lambda, \, 1}(k)|^2 dk
=  C \, \frac{2}{3}\cdot  \frac{1}{(2\pi)^3}\cdot 4\pi \Lambda
$$
which leads to $C= {3}{\pi^2}\Lambda^{-1} $. We also have
$$
\sum_\lambda 
\sum_j \int_{|k| < \Lambda}  |k| \,
|\widehat{v}_{\lambda,j} (k)|^2 dk  \  \int_{\R^3} w(y)dy 
= 2\cdot (2\pi)^{-3} \cdot
(4\pi/2)  \Lambda^2 C  \  ,
$$ 
so that  (\ref{j.4}) becomes
\begin{equation}\label{j.12}
\boxed{\quad
H_f  \geq  \frac{3\pi}{8}\Lambda^{-1}A^\Lambda(x)^2 -\frac{3}{ 4}\Lambda
\quad}
\end{equation}

\section{Appendix: Spectral properties of the Dirac operators}
\label{sec:commute}

In this appendix we sketch a proof of the fact that the operators $D_i(A)$
commute in the sense that all their spectral projections commute.
First we start with some remarks concerning the self-adjointness of
\begin{equation}
D_i(A)= {{\mbox{\boldmath$\alpha$}_i} }\cdot \left( -i \nabla_i + \sqrt{\alpha} 
A(x_i) \right) +m \beta \ .
\end{equation}
The subscript after ${{\mbox{\boldmath$\alpha$} } }$ is a reminder that
the matrix acts on the spinor associated with the $i$-th particle.  It is
not easy to characterize the domain for this operator, but it is certainly
defined and symmetric on ${\cal H}^0_N := C^\infty_c(\RR) \bigotimes
{\mathcal F}_{\mathrm {finite}}$, where ${\mathcal F}_{\mathrm {finite}}$
denotes the vectors in Fock space with finitely many photons. This is a
dense subset of ${\cal H}_N$.  We shall show that $D_i(A)$ is essentially
self-adjoint on ${\cal H}^0_N$. To prove this we resort to a version of
Nelson's commutator theorem given in \cite{reedsimon2}, Theorem X.37.

Define the operator 
\begin{equation}
\nu = { 1 + \sum_{i=1}^N (-\Delta_i + m^2) + \Lambda H_f + \Lambda^2} \ ,
\end{equation}
Observe that  
$\sum_{i=1}^N (-\Delta_i + m^2)$ acts as a multiplication operator on 
Fourier space and $H_f$ acts on the $n$ photon component 
$\Phi_n(x_1, \dots, x_N;\tau_1, \dots,
\tau_N;k_1, \dots, k_n, \lambda_1, \dots, \lambda_n)$ by 
\hfill\break multiplication with
$\sum_{i=1}^n \omega(k_i)$. The domain ${\cal H}^0_N$ is a domain of
essential self-adjointness for $\nu $. 
Certainly, $\nu \geq 1$ as an operator.

We shall show that there exists a constant $c$ such that for all 
$\Psi \in {\cal H}^0_N$, 
\begin{equation} \label{relbound}
\Vert D_i(A) \Psi \Vert \leq c \Vert \nu  \Psi \Vert \ .
\end{equation}
Certainly,
\begin{equation}
\Vert D_i(A) \Psi \Vert \leq 
\left(\Psi, (-\Delta_i+m^2) \Psi \right)^{1/2} +(\Psi, A(x_i)^2 \Psi)^{1/2}
\end{equation}
and by Example 4 in Appendix \ref{sec:appendix}, 
\begin{equation}
A(x)^2 \leq \frac{8}{3 \pi} \Lambda H_f + \frac{2}{\pi}\Lambda^2    \ .
\end{equation}
The estimate
\begin{equation}
\Vert D_i(A) \Psi \Vert  \leq c \Vert \nu  \Psi \Vert 
\end{equation}
follows easily from this.

Next, we show that there exists a constant $d$ such that 
for all $\Psi \in {\cal H}^0_N$
\begin{equation}
|\left( D_i(A) \Psi, \nu  \Psi \right) - \left( \nu  \Psi,D_i(A) \Psi \right)|
\leq d \Vert \nu^{1/2} \Psi \Vert ^2 \ . \label{commuting}
\end{equation}
Since $-i {{\mbox{\boldmath$\alpha$}_i} }\cdot \nabla_i  +m \beta$ commutes with
$\nu $ when applied to vectors in ${\cal H}^0_N$, the above estimate reduces
to
\begin{equation}
|\left( {{\mbox{\boldmath$\alpha$}_i} } \cdot A(x_i) \Psi, \nu \Psi \right) 
- \left( \nu \Psi, {{\mbox{\boldmath$\alpha$}_i} } \cdot A(x_i) \Psi \right)|
\leq d \Vert \nu^{1/2} \Psi \Vert ^2 \ ,
\end{equation}
where we have dropped the fine structure constant. Since $\nu $ as well as 
$A(x)$ preserve ${\cal H}^0_N$ and are symmetric, we can rewrite the 
above inequality as
\begin{equation}
|\left(\Psi, \left[{{\mbox{\boldmath$\alpha$}_i} } \cdot A(x_i) ,\nu \right]
 \Psi \right) | \leq d \Vert \nu^{1/2} \Psi \Vert ^2 \ .
\end{equation}

The commutator is the sum of
\begin{equation}
\left[{{\mbox{\boldmath$\alpha$}_i} } \cdot A(x), -\Delta \right]
=i\sum_{i=1}^3 {{\mbox{\boldmath$\alpha$}_i} } \cdot
\left( \partial_j A(x) \partial_j + \partial_j\partial_j A(x) \right)=: X \ ,
\end{equation}
and the operator
\begin{equation}
-i\Lambda {{\mbox{\boldmath$\alpha$}_i} } \cdot E(x_i) =\left[{{\mbox{\boldmath$\alpha$}_i} } \cdot A(x_i), 
\Lambda H_f \right] \ ,
\end{equation}
where $E(x)$ is the electric field (\ref{e}).
By Schwarz's inequality, 
\begin{equation}
|\left( \Psi, X \Psi \right)| \leq \left( \Psi, (\nabla A)^2 \Psi) \right)
+ \left( \Psi, -\Delta \Psi) \right) \ , \label{comone}
\end{equation}
and
\begin{equation}
\Lambda |\left(\Psi, E(x) \Psi \right)| \leq \Lambda \left( \Vert \Psi \Vert ^2 
+ (\Psi, E(x)^2 \Psi) \right) \ . \label{comtwo}
\end{equation}
By Example 3 in Appendix \ref{sec:appendix}, it follows that as quadratic forms
\begin{eqnarray}
& E(x)^2 \leq \frac{8}{9 \pi} \Lambda^3 H_f + \frac{1}{\pi}\Lambda^4 \\
& (\nabla A)(x)^2 \leq \frac{8}{9 \pi} \Lambda^3 H_f + \frac{1}{\pi} \Lambda^4 
\label{nabA} \ .
\end{eqnarray}
The last inequality does not appear in Appendix \ref{sec:appendix}
exactly as stated, 
but it can be
derived in precisely the same fashion as the one for the magnetic field
displayed there. The estimates  (\ref{comone})-(\ref{nabA})
yield (provided $\Lambda \geq 1$)
\begin{eqnarray}
& |\left( D_i(A) \Psi, \nu \Psi \right) - \left( \nu \Psi,D_i(A) 
\Psi \right)| \\
& \leq C \left(\Psi, (\Lambda^3 H_f + \Lambda^4) \Psi\right) 
+\left (\Psi, (-\Delta
+\Lambda) \Psi \right) \\
& \leq C \Lambda^2 \left(\Psi, \nu \Psi \right) \ ,
\end{eqnarray}
for some constant $C$, which is the desired estimate. 

Thus, the operator $D_i(A)$ is essentially self-adjoint on ${\cal H}^0_N$. This 
operator,
being a sum of two self-adjoint operators, 
$U_i:=-i {{\mbox{\boldmath$\alpha$}_i} }\cdot \nabla_i +m \beta$
and $V_i:={{\mbox{\boldmath$\alpha$}_i} }\cdot \sqrt{\alpha}  A(x_i) $, is 
naturally
defined on ${\cal D}(U_i) \cap {\cal D}(V_i)$ and is symmetric there. 
Since, ${\cal H}^0_N \subset {\cal D}(U_i) \cap {\cal D}(V_i)$ we also
know that $D_i(A)$ is essentially self-adjoint on ${\cal D}(U_i) \cap {\cal 
D}(V_i)$. Thus, by 
Theorem VIII.31 in \cite{reedsimon1} the Trotter product formula is valid, i.e.
\begin{equation}
e^{itD_i(A)} = s-\lim_{m \to \infty} T_i(t/m)^m  \ ,
\end{equation}
where
\begin{equation}
T_i(t/m) :=  e^{i(t/m)U_i} e^{i(t/m) V_i}  \ .
\end{equation}
Certainly, the operator $e^{itU_i}$ commutes with $e^{isU_j}$ and $e^{isV_j}$,
and likewise $e^{itV_i}$ commutes with $e^{isU_j}$ and $e^{isV_j}$ for all
$j \not= i$, and hence $T_i(t/m)^m$ commutes with $T_j(s/n)^n$ for all $s,t, m$ 
and $n$.

We shall use this to show the following
\begin{lm} \label{unitary}
For any two real numbers $s$ and $t$ the unitary groups $e^{itD_i(A)}$ and
$e^{isD_j(A)}$ commute. Moreover, this implies that the spectral projections
associated with $D_i(A)$ and $D_j(A)$ commute.
\end{lm}
\begin{proof}
 For $\Psi \in {\cal H}_N$
\begin{eqnarray}
&\Vert e^{itD_i(A)}e^{isD_j(A)} \Psi - e^{isD_j(A)}e^{itD_i(A)} \Psi \Vert \\
& =\lim_{m \to \infty} \Vert T_i(t/m)^me^{isD_j(A)} \Psi - 
e^{isD_j(A)}T_i(t/m)^m \Psi \Vert \\
& =\lim_{m \to \infty} \lim_{n \to \infty} 
\Vert T_i(t/m)^mT_j(s/n)^n \Psi - T_j(s/n)^nT_i(t/m)^m \Psi \Vert  = 0\ .
\end{eqnarray}
The statement about the spectral projections follows from Theorem VIII.13
in \cite{reedsimon1}.
\end{proof}

\section{Appendix: Projections and symmetries}
\label{projections}
The difficulty in defining the physical space ${\mathcal
H}^{\mathrm{phys}}_N$ comes from the fact that the projection onto
the positive energy subspace acts also on the Fock space. This is in
contrast to \cite{LSS} where no such problem arises.  There the action
of the permutation group obviously commuted with the projection onto the
positive energy subspace. In our more general setting the commutation
is still true but an explanation is needed, which we try to give with
a minimal amount of formality.

First, consider the one particle space ${\mathcal{H}}_1 = L^2(\R^3; \C^4)
\bigotimes {\mathcal{F}}$. The Dirac operator, as shown in Appendix
\ref{sec:commute}, is a self-adjoint operator on ${\mathcal{H}}_1$ and we
denote its projections onto the positive and negative energy subspace by
$P^+$ and $P^-$. Note that $P^+ + P^-$ is the identity. As explained in
Section \ref{nopair}, the two projections are unitarily equivalent via
\begin{equation}\label{pu1}
P^- = U^* P^+ U = -U P^+ U
\end{equation}
where, as in (\ref{you}),
\begin{equation} \label{pu2}
U= \begin{pmatrix} 0 & {\mathrm{I}} \\ -{\mathrm{I}} & 0 
\end{pmatrix} \ .
\end{equation}
The projection onto the positive energy subspace associated with the
Dirac operator $D_i(A)$  is defined in the following fashion. Consider
the vector $\Psi$ as in (\ref{psi}) and fix $N-1$ $x$'s and $\tau$'s,
namely all those except $x_i$ and $\tau_i$. For almost every such choice
(with respect to Lebesgue measure) the vector $\Psi$ defines a vector
in ${\mathcal{H}}_1$. We know how the $P^{\pm}$ act on such a vector
and the extension to ${\mathcal{H}}_N$ we denote by $P^{\pm}_i \Psi$.
It was shown in Appendix \ref{sec:commute} that these spectral projectors
commute with each other.

Other interesting operators on ${\mathcal{H}}_N$ are the permutations.
A permutation ${\rm Per}_{1,2}$, for example, just exchanges the electron
labels $1$ and $2$. From what has been explained above we have the formula
\begin{equation}
{\rm Per}_{1,2} P^{\pm}_1 = P^{\pm}_2 {\rm Per}_{1,2} \ .
\end{equation}
An immediate consequence is that $P^{\pm} := \Pi_{i=1}^N P^{\pm}_i$ commutes
with permutations.

{}From this it follows that (\ref{physhil}) can be rewritten as
\begin{equation}\label{equiv}
{\mathcal{H}}^{\mathrm{phys}}_N := {\mathcal{A}}P^{+}{\mathcal{H}}_N
=P^{+}{\mathcal{A}}{\mathcal{H}}_N \ .
\end{equation}

We now address the question whether 
${\mathcal{H}}^{\mathrm{phys}}_N$ is trivial 
or not.
Denote by ${\mathcal{K}}$ the subspace of ${\mathcal{H}}_N$ which consists of 
antisymmetric
vectors $\Psi$ with the property that $U_j \Psi = i \Psi$ for $j=1, \dots, N$. 
The operators
$U_j$ are defined in (\ref{you}). Certainly each $U_j$ has eigenvalues $i$ and 
$-i$.
The space ${\mathcal{K}}$ is certainly infinite dimensional. It contains, e.g., 
determinantal
vectors in $L^2(\R^3;\C^4)$ tensor the photon vacuum.

\begin{lm}[${\mathcal{H}}^{\mathrm{phys}_N}$ is large]
The space ${\mathcal{H}}^{\mathrm{phys}}_N$ is infinite dimensional.
\end{lm}

\begin{proof}
We shall show that $2^{N/2} P^+$ is an isometry from ${\mathcal{K}}$ into 
${\mathcal{H}}^{\mathrm{phys}}_N$. Let $I$ be a subset of the integers 
$\{1, \dots, N\}$ and let $J$ be its complement. Let 
 \begin{equation}
P_{I} = \Pi_{i \in I} P^-_i \Pi_{j \in J} P^+_j \ .
\end{equation}
Note that $\sum_I P_{I} = {\rm identity}$.
Note also that 
\begin{equation}
P_{I} = (-)^{|I|} U_I \Pi_{i \in I} P^+_i U_I\Pi_{j \in J} P^+_j
\end{equation}
which implies that $\Vert P_{I}\Psi  \Vert = \Vert P^+ \Psi \Vert$. 
This shows in particular that
\begin{equation}
\Vert P^+ \Psi \Vert^2 = 2^{-N}  \Vert \Psi \Vert^2 \ ,
\end{equation}
which proves the isometry.
\end{proof}

Since we always consider the symmetric operator $H^{\mathrm{phys}}_N$ 
in the sense of quadratic forms,
it is necessary to construct a domain, ${\mathcal{Q}}_N$ that 
is dense in ${\mathcal{H}}^{\mathrm{phys}}_N$ and on which every
term in $H^{\mathrm{phys}}_N$ has a finite expectation value. Once it is
shown that the quadratic form associated with $H^{\mathrm{phys}}_N$
is bounded below, it is closable and its closure defines a self-adjoint
operator, the Friedrich's extension of $H^{\mathrm{phys}}_N$.

We first start with a technical lemma that will allow us to approximate
vectors in ${\mathcal{H}}^{\mathrm{phys}}_N$.

\begin{lm} \label{eff}
For any $\widehat{f}$ with 
\begin{equation}
C_f := \int_{-\infty}^{\infty} |\widehat{f}(t)|(1+|t|) {\rm d} t < \infty
\end{equation}
and 
\begin{equation}
f(D(A)) = \int e^{-itD(A)} \widehat{f}(t) {\rm d}t
\end{equation}
we have that
\begin{equation}
\Vert \sqrt{1+H_f} f(D(A)) \Psi \Vert  \leq 
\max \{\sqrt{1+9 \Lambda/8},\sqrt{8/9\pi} \Lambda ^{3/2}\}
 C_f\Vert \sqrt{1+H_f} \Psi \Vert \ ,
\end{equation}
for all $\Psi \in {\mathcal{H}}_1$. 
\end{lm}

\begin{proof} We shall assume that $\Psi$ is normalized.
 Since
 \begin{equation}
 \Vert \sqrt{1+H_f} f(D(A)) \Psi \Vert 
  \leq \int |\widehat{f}(t)|\Vert \sqrt{1+H_f} e^{-itD(A)} 
 \Psi \Vert {\rm d}t \ ,
 \end{equation}
 it suffices to prove the estimate
 \begin{equation}
 K(t) := \Vert \sqrt{1+H_f}\ e^{-itD(A)} \Psi \Vert \leq
 C(1+|t|)\Vert \sqrt{1+H_f} \Psi \Vert\ .
 \end{equation}
 
 A simple calculation yields
 \begin{equation}
 \frac{{\rm d}}{{\rm d}t} K^2(t) =
 \left(e^{-itD(A)} \Psi, i[H_f, D(A)] e^{-itD(A)} \Psi \right)
 = -\left(e^{-itD(A)} \Psi, \alpha \cdot E(x) e^{-itD(A)} \Psi \right) \ .
 \end{equation}
 Here $E(x)$ is the electric field
 \begin{equation}
 E(x) = \frac{i}{2\pi} \sum_{i=1}^2 \int_{|k| \leq \Lambda} {\mathrm d} k
  \varepsilon_{\lambda}(k) \sqrt{\omega(k)} \left(e^{ik \cdot x} 
 a_{\lambda}(k) - e^{-ik \cdot x}a_{\lambda}^*(k) \right) \ .
 \end{equation} 
 By Schwarz's inequality
 \begin{equation}
 \frac{{\rm d}}{{\rm d}t} K^2(t) \leq 
 \left(e^{-itD(A)} \Psi, E(x)^2 e^{-itD(A)} \Psi \right)^{1/2} \ .
 \end{equation}
 By Example 3 in Section \ref{sec:appendix}
 \begin{equation}
 E(x)^2 \leq \frac{8}{9\pi} \Lambda^3 H_f + \frac{1}{\pi} \Lambda^4
 \end{equation}
 and hence $K^2(t)$ satisfies  the differential inequality
 \begin{equation}
 \frac{{\rm d}}{{\rm d}t} K^2(t)  \leq
 \left( A K^2(t) + B \right)^{1/2} \ ,
 \end{equation}
 where $A= \frac{8}{9\pi} \Lambda^3$ and $B=\frac{1}{\pi} \Lambda^4$.
 This can be readily solved (using $K(t) \geq 1$) to yield the estimate
 \begin{equation}
 K(t) \leq (1 + {B}/{A})^{1/2} K(0) + \sqrt{A}t \ .
 \end{equation}
 Thus
 \begin{equation}
 \Vert \sqrt{1+H_f} e^{-itD(A)} \Psi \Vert
 \leq C(1+|t|)\Vert \sqrt{1+H_f} \Psi \Vert \ ,
 \end{equation}
 where $C$ is the maximum of $ (1 + {B}/{A})^{1/2}$ and $\sqrt A$. 

\end{proof}

Next we consider a sequence of functions 
$f_n \in C^{\infty}_c((0, \infty))$ everywhere less or equal to
$1$, such that
$f_n$ is identically equals to $1$ on the interval $[1/n, n]$.
Clearly, as $n \to \infty$, $\Pi_{i=1}^Nf_n(D_i(A)) \to P^+$
strongly in ${\mathcal{H}}_N$ and hence
${\mathcal{A}}\Pi_{i=1}^Nf_n(D_i(A)) \to  {\mathcal{A}}P^+$
strongly in ${\mathcal{H}}_N$. We denote the range of 
${\mathcal{A}}\Pi_{i=1}^Nf_n(D_i(A))$ restricted to the subspace of 
${\mathcal{H}}_N$ consisting of states with {\it finite field energy 
expectation} by
${\mathcal{Q}}^n_N$. Finally we define the domain
${\mathcal{Q}}_N = \cup_{n=1}^{\infty}{\mathcal{Q}}^n_N$.
Together with Lemma \ref{eff} we have the following Corollary. 

\begin{cl}\label{field} The domain ${\mathcal{Q}}_N$ is dense in 
${\mathcal{H}}^{\mathrm{phys}}_N$. Moreover for any vector
$\Psi \in {\mathcal{Q}}_N$ the field energy $H_f$ has finite 
expectation value.
\end{cl}
\begin{proof}
Simply note that the functions $f_n$ have a rapidly decaying
Fourier transform for each $n$. Therefore, by Lemma \ref{eff}
the field energy has a finite expectation value for any vector $\Psi
\in {\mathcal{Q}}^n_N$. Note, as before, the antisymmetrization 
operator ${\mathcal{A}}$ commutes with $\Pi_{i=1}^Nf_n(D_i(A))$.
Thus, the field energy has finite expectation value for any
$\Psi \in {\mathcal{Q}}_N$. The density of ${\mathcal{Q}}_N$
in ${\mathcal{H}}^{\mathrm{phys}}_N$ was shown before.
\end{proof}

Now we are ready to state the main lemma of this section.
\begin{lm}
For every $\Psi \in {\mathcal{Q}}_N$, the Dirac operators $D_i(A)$,
the Coulomb potential $V_c$ and the field energy $H_f$ have finite
expectation values. Thus, $H^{\mathrm{phys}}_N$ is defined 
 as a quadratic form on ${\mathcal{Q}}_N$ which is dense in
${\mathcal{H}}^{\mathrm{phys}}_N$.
\end{lm}
\begin{proof}
The operators $D_i(A)^2$ have finite expectation values
on ${\mathcal{Q}}_N$. They are of the form $\widehat{T}^P(A)
=\left[(p+\sqrt{\alpha}A(x))^2 + \sqrt{\alpha}\sigma \cdot B(x)
\right] \otimes I_2$ where $I_2$ is the $2 \times 2$ identity. By
(\ref{j.11}) the magnetic field is bounded by the field energy and hence
has finite expectation value on the domain ${\mathcal{Q}}_N$. Thus,
$(p_i+\sqrt{\alpha}A(x_i))^2 \otimes I_2$  also has finite expectation
value on ${\mathcal{Q}}_N$ for $i=1, \dots, N$, and hence the
Coulomb potential $V_c$, which is relatively bounded with respect to
$\sum_{i=1}^N (p_i+\sqrt{\alpha}A(x_i))^2$ has finite expectation values
on ${\mathcal{Q}}_N$.  
\end{proof}

\section{Appendix: Various forms of instability}
\label{sec:instability}

In the introduction we talked about the need of using the positive
spectral subspace of the Dirac operator $D(A)$, which includes the
magnetic vector potential; this led to all sorts of complications in the
analysis leading to our main stability Theorem \ref{mainthm}. In this
section we show that various models in which an electron is defined,
instead, by the positive spectral subspace of the {\it free} Dirac
operator $D(0)$ are unstable. In the case of a classical magnetic field
such an analysis was carried out in \cite{LSS} and greatly simplified
in \cite{GT}. Also, in \cite{GT} the stability analysis was carried
out for a quantized radiation field without a cutoff. In what follows,
we rely mostly on the work in \cite{GT}. We also show that the 
$D(A)$ choice is unstable if $Z\alpha$ or $\alpha$ is too large --- as 
expected.

All the results about stability and instability are summarized in the two
tables in Section \ref{intro}.  We remind the reader that instability
of the first kind means that the Hamiltonian is unbounded below, while
instability of the second kind means that it is bounded below but not
by a constant times $N+K$.

\subsection{Instability without Coulomb potential (free Dirac operator)}
Already the free problem, i.e., without Coulomb interactions, shows signs
of instability. The Hamiltonian is given by
\begin{equation}
H^{\mathrm{q}}_N = \sum_{j=1}^N D_j(A) + H_f \ .
\end{equation}
If the field is classical $H_f$  has to be replaced by
$(1/8\pi) \int_{\R^3} |B(x)|^2 {\rm d} x$ as in (\ref{fieldenergy}).

We consider first the case where the magnetic vector potential is classical.
In particular the Hilbert space ${\mathcal{H}}^{\mathrm{free}}_N$
is the antisymmetric tensor product of
$N$ copies of $P^+L^2(\R^3;\C^4)$, i.e., the part of $L^2(\R^3;\C^4)$
that is in the positive spectral subspace of the {\it free} Dirac operator.
Note that there is no Fock space in this case.
In \cite{GT} Theorems 1 and 3 the authors construct, for any $N$,
a trial Slater determinant
$\psi$ in ${\mathcal{H}}^{\mathrm{free}}_N$, and a classical field $A$ 
so that the energy is bounded above by
\begin{equation} \label{gtone}
\left(\psi,H^{\mathrm{classical}}_N \psi \right) =: {\mathcal E}(\psi, A) 
\leq aN^{4/3} - \alpha b N^2 \ ,
\end{equation}
where  $a$ and $b$ are constants independent of $N$.  The scaling
\begin{equation}
\psi \to \psi_{\mu}\ ,\ \ {\mathrm {and}} \ \ A \to A^{\mu}
\end{equation}
where
\begin{equation}
\psi_{\mu}(x_1, \dots , x_N; \tau_1, \dots , \tau_N) =
\mu^{3N/2} \psi(\mu x_1, \dots ,\mu  x_N; \tau_1, \dots , \tau_N)
\end{equation}
and
\begin{equation}
A^{\mu}(x) = \mu A(\mu x)
\end{equation}
can be used to get the upper bound
\begin{equation} \label{gtformula}
{\mathcal E}(\psi_{\mu}, A^{\mu}) \leq \mu 
\left(AN^{1/3} - \alpha B N^2 \right) \ .
\end{equation}
Thus, by choosing $N > a \alpha ^{-3/2}/b$, the ground state energy
is negative and can be driven to $-\infty$ by letting $\mu\to \infty$,
i.e., {\it stability of the first kind is violated.}

Using coherent states, it was shown in \cite{GT} that this same result
extends to the problem with a quantized magnetic vector potential
without ultraviolet cutoff. 

If $A(x)$ carries an ultraviolet cutoff the $\mu$ scaling argument cannot
be applied. The energy, however, is not bounded below by ${\rm const.}
\times N$, as we see from (\ref{gtone}), and hence {\it stability of
the second kind} is also violated --- and this no matter how small   
$\alpha$ may be and whether the magnetic vector
potential is quantized or not. The reader might wonder how to construct
an $A(x)$ that satisfies the conditions in \cite{GT} and, at the same
time has an ultraviolet cut off $\Lambda$. Remark 1 on page 1782 of
\cite{GT} explains that almost any cutoff $A(x)$ will suffice for the
purpose.

Notice, that when we use the positive spectral subspace of $D(A)$ instead,
the stability of the  problem without Coulomb potential is completely
trivial, since the Hamiltonian is positive, by definition.

\subsection{Instability with Coulomb potential (free Dirac operator)}  
Adding the Coulomb potential complicates the analysis owing to the 
repulsion between the electrons which is present even if there are 
no nuclei. To some extent this positive 
energy is balanced by the electron-nuclei attraction
if sufficiently many nuclei with sufficiently
strong charges are present. It is shown in \cite{LSS} that if
\begin{equation}
\sum_{j=1}^KZ_j \geq ({\mathrm {const.}}) \alpha^{-3/2} \ \ {\rm and} 
\ \ \sum_{j=1}^KZ_j^2 \geq 2  \ , \label{instabcond}
\end{equation}
then the positions of the nuclei can be chosen such that the total Coulomb
energy is negative.
Thus, if in addition, $N\alpha^{3/2}$ is sufficiently large, 
{\it stability of the first kind} does not hold for classical magnetic vector
potentials as well as for a quantized magnetic vector potential (without
ultraviolet cutoff) --- no matter how small $\alpha$ may be.
 
The situation is more complicated when the field carries an ultraviolet
cutoff. The main reason is that the field variable is no longer an active 
participant for driving the energy towards minus infinity, but it is an active
participant in destroying stability of the second kind. We have
\begin{lm} \label{lemmm}
Let $\alpha > 0$ and assume that (\ref{instabcond}) holds. Then the system
using the projection onto the positive subspace of the free Dirac
operator $D(0)$
is unstable of the second kind, even with an ultraviolet cutoff.
\end{lm}
\begin{proof}
The lemma follows immediately from (\ref{gtformula}) and
(\ref{instabcond}) together with the observation in \cite{GT} on how to
use coherent states to
carry these results over to the  quantized field case.
\end{proof}

{\emph {Lemma \ref{lemmm}  is the main reason that the restriction to
the positive spectral subspace of $D(0)$ is inadequate for a model of
matter interacting with radiation.}}

The main result of this paper is the stability of the second kind for the
system where the {\it positive subspace} of the Dirac operator $D(A)$
is used.  This result holds provided that  $\max_j Z_j \alpha$ and 
$\alpha$ is sufficiently small. Our final Lemma \ref{fourpi} shows that
that the two conditions $\max_j Z_j \alpha$ small and $\alpha$ 
small are in fact necessary. 
It suffices to show this for the
case of one electron interacting with $K$ nuclei, each having charge $Z$.
We have to assume that the electron mass is strictly positive, but we suspect
that this assumption is technical and not needed.

\begin{lm}\label{fourpi}
Assume that $Z\alpha > 4/\pi $ and $m>0$. Then the one-electron Hamiltonian
\begin{equation}
H^{\mathrm{phys}}_1=P^+\left( D(A)-Z\alpha\sum_{j=1}^K\frac{1}{|x-R_j|} 
+Z^2 \alpha \sum_{i<j}\frac{1}{|R_i-R_j|}+ H_f \right)P^+
\end{equation}
is unbounded below. Moreover, there is a number $\alpha_c$ 
such that for any fixed $\alpha >\alpha_c$ and any fixed $Z>0$,
there exists $K$ sufficiently large so that this Hamiltonian is unbounded 
below. 
\end{lm}

For the classical instead of the quantized $A$ field see
\cite{Evansetal1996, HS, BE, Tix1, Tix2}.

\begin{proof}
The idea is to reduce this problem to the relativistic one without spin and
without radiation field.
Set $D_+ =P^+D(A)$ and $D_- =-P^-D(A)$, so that $D(A)=D_+ - D_-$ and
$|D(A)|=D_+ + D_-$. As a trial function we pick $\psi = g \otimes |0 \rangle$,
where $g$ is a spinor that satisfies $Ug = ig$, recalling the definition 
of $U$ 
from Appendix \ref{projections} (\ref{pu1}, \ref{pu2}), and where
$|0 \rangle$ is the photon vacuum. A straightforward calculation (which
repeatedly uses
Schwarz's inequality and the facts that $P^- = U^*P^+U = -UP^+U$, and
hence $P^-\psi = -i\,  UP^+\psi$) shows that
\allowdisplaybreaks
\begin{eqnarray}
\left( \psi, H^{\mathrm{phys}}_1 \psi \right) 
&=&\left(P^+\psi, \left[D_+ -Z\alpha\sum_{j=1}^K\frac{1}{|x-R_j|} 
+Z^2 \alpha \sum_{i<j}\frac{1}{|R_i-R_j|}+H_f \right] P^+\psi \right) 
\notag\\
&=&\frac{1}{2} \left(P^+\psi, \left[D_+ -Z\alpha\sum_{j=1}^K\frac{1}{|x-R_j|} 
+Z^2 \alpha \sum_{i<j}\frac{1}{|R_i-R_j|}\right] P^+\psi \right)  \notag\\
&+&\frac{1}{2}\left(P^-\psi, \left[D_- -Z\alpha\sum_{j=1}^K\frac{1}{|x-R_j|} 
+Z^2 \alpha \sum_{i<j}\frac{1}{|R_i-R_j|}\right] P^-\psi \right) \notag \\
&+& \left(\psi,P^+H_fP^+ \psi\right) \notag \\
&\leq &\frac{1}{2} \left(P^+\psi, \left[D_+ -\frac{Z\alpha}{2}
\sum_{j=1}^K\frac{1}{|x-R_j|} 
+Z^2 \alpha \sum_{i<j}\frac{1}{|R_i-R_j|}\right] 
P^+\psi \right) \notag \\
&+& \frac{1}{2}\left(P^-\psi, \left[D_- -\frac{Z\alpha}{2}
\sum_{j=1}^K\frac{1}{|x-R_j|} 
+Z^2 \alpha \sum_{i<j}\frac{1}{|R_i-R_j|}\right] P^-\psi \right)\notag \notag\\
&-& {\rm Re}\left(P^+\psi, \frac{Z\alpha}{2}\sum_{j=1}^K\frac{1}{|x-R_j|}
P^-\psi  \right) +\left(\psi,P^+H_fP^+ \psi\right) \notag \\
&\leq&\frac{1}{2} \left(\psi, \left[|D(A)| -\frac{Z\alpha}{2}
\sum_{j=1}^K\frac{1}{|x-R_j|} 
+Z^2 \alpha \sum_{i<j}\frac{1}{|R_i-R_j|}
 \right] \psi \right) + \left(\psi,P^+H_fP^+ \psi\right) \notag \ .
\end{eqnarray}
The lemma will be  proved by showing that the first term
in the last expression can be made as negative as we like while the field energy
term is uniformly bounded.
Note, as in (\ref{dsquare})
 that $|D(A)| = \sqrt{\widehat{T}^P(A) + m^2}$. The operator inequality 
\begin{equation}
(p+\sqrt{\alpha} A(x))^2 \leq (1+ \varepsilon) p^2 + 
(1+ \frac{1}{\varepsilon})\alpha A(x)^2
\end{equation} 
follows easily from Schwarz's inequality, 
for any $\varepsilon>0$. From (\ref{j.12})  we have that
\begin{equation}
A(x)^2 \leq \frac{8}{3 \pi} \Lambda H_f + \frac{2}{\pi} \Lambda^2
\end{equation}
and from (\ref{j.11}) we have that
\begin{equation}
B(x)^2 \leq \frac{8}{9\pi}\Lambda^3 H_f + \frac{1}{\pi} \Lambda^4 \ .
\end{equation}
Using the operator monotonicity of the square root it follows that
\begin{equation}
\sqrt{T^P(A) + m^2} \leq \sqrt{(1+ \varepsilon) p^2 + X_{\varepsilon}
\Lambda H_f + Y_{\varepsilon}\Lambda^2+m^2}
\end{equation}
Where $X_{\varepsilon}$ and $Y_{\varepsilon}$ are constants that tend to 
infinity as $\varepsilon$
tends to zero.

Thus, recalling  that $\psi= g\otimes |0 \rangle$
\begin{equation}
\left(\psi, |D(A)| \psi \right) \leq \left(g,\sqrt{(1+ \varepsilon) p^2 
+ Y_{\varepsilon}\Lambda^2} \ g \right) \ .
\end{equation}

The remaining task is to analyze the quadratic form
\begin{equation}
\left(g,\left[\sqrt{(1+ \varepsilon) p^2 
+ Y_{\varepsilon}\Lambda^2+m^2} -\frac{Z\alpha}{2}\sum_{j=1}^K\frac{1}
{|x-R_j|} 
+Z^2 \alpha \sum_{i<j}\frac{1}{|R_i-R_j|} \right]g \right)  \ .
\end{equation}
For any fixed $\varepsilon$, $\Lambda$ and $m$ the terms 
$ Y_{\varepsilon}\Lambda^2+m^2$
can be scaled away, and this leads to the quadratic form
\begin{equation}
\left(g,\left[\sqrt{(1+ \varepsilon) p^2 } -\frac{Z\alpha}{2}
\sum_{j=1}^K\frac{1}{|x-R_j|} 
+Z^2 \alpha \sum_{i<j}\frac{1}{|R_i-R_j|} \right]g \right)  \ ,
\end{equation}
which has been analyzed in detail.
Kato \cite{Kato} showed that instability of the first kind occurs if
$Z \alpha/2\sqrt{1+ \varepsilon} >2 /\pi$ which yields our first stated 
condition
for instability. Later on, it was shown in \cite{DL} that there exists
$\alpha_c$ so that for $\alpha > \alpha_c$ and for any $Z>0$ there exists
$K$ so that an instability of the first kind occurs.  
See also \cite{LY}, Theorem 3.

Next we address the field energy term and it is here where the assumption 
about the positive mass comes in. The projection $P^+$ can be written as
$P^+=(1/2)(I+D(A)/|D(A)|)$ and, since $\psi$ is proportional to the vacuum,
\begin{equation}
\left(\psi, P^+ H_f P^+ \psi \right) = \frac{1}{4} 
\left(\psi, \frac{D(A)}{|D(A)|}H_f \frac{D(A)}{|D(A)|} \psi \right)  \notag 
= \sum_{\lambda} \int \omega(k) \left\Vert \left[a_{\lambda}(k),\frac{D(A)}{|D(A)|}\right] 
\psi \right\Vert^2 {\rm d}k \ . \label{fieldcommutator}
\end{equation}
With the help of the expression 
\begin{equation}
\frac{D(A)}{|D(A)|}=\frac{1}{\pi} \int_0^{\infty} \frac{D(A)}{t+D(A)^2} 
\frac{1}{\sqrt t}{\rm d}t \ ,
\end{equation}
and the fact that $(t+D(A)^2)^{-1} =\frac{1}{2} (D(A)-i\sqrt t)^{-1}+
\frac{1}{2} (D(A)+i\sqrt t)^{-1}$,
the commutator in the last expression of (\ref{fieldcommutator}) can
be written as
\begin{eqnarray}
\left[a_{\lambda}(k),\frac{D(A)}{|D(A)|}\right] =
&-&\frac{1}{2\pi}\int_0^{\infty} \frac{1}{D(A)-i\sqrt t}
\left[a_{\lambda}(k),D(A)\right]\frac{1}{D(A)-i\sqrt t} \frac{1}{\sqrt t}{\rm d}t \notag\\
&-&\frac{1}{2\pi}\int_0^{\infty}\frac{1}{D(A)+i\sqrt t}\left[a_{\lambda}(k),D(A)\right]
\frac{1}{D(A)+i\sqrt t} \frac{1}{\sqrt t} {\rm d}t \ .
\end{eqnarray}
and 
$$
\left[a_{\lambda}(k),D(A)\right] = 
\frac{\varepsilon_{\lambda}(k)}{\sqrt{\omega(k)}}e^{-ik\cdot x}\chi_{\Lambda}(k) 
$$
where $\chi_{\Lambda}(k)$ denotes the characteristic function of the  ball in $k$ space 
that has radius $\Lambda$  and is centered
at the origin.
Hence,
\begin{equation}
\left\Vert \left[a_{\lambda}(k),\frac{D(A)}{|D(A)|}\right] 
\psi \right\Vert \leq \frac{1}{\sqrt{\omega(k)}}\frac{1}{\pi} \int_0^{\infty}
\frac{1}{m^2+t}\frac{1}{\sqrt t} {\rm d}t \Vert \psi \Vert \chi_{\Lambda}(k)
= \frac{1}{\sqrt{\omega(k)}m}\Vert \psi \Vert \chi_{\Lambda}(k) \ ,
\end{equation} 
and the estimate
\begin{equation}
\left(\psi, P^+ H_f P^+ \psi \right) \leq \frac{4 \pi \Lambda^3}{3m^2} \Vert \psi \Vert^2
\end{equation}
follows.
\end{proof}


\begin{thebibliography}{99}

\bibitem{BE}
A.~A. Balinsky and W.~D. Evans,
\newblock \textit{Stability of one-electron molecules in the Brown-Ravenhall
model,}
\newblock Commun. Math. Phys. {\bf 202}, 481-500 (1999).


\bibitem {BKS} M.S. Birman, L.S. Koplienko and M.Z. Solomyak,
\textit{Estimates for the spectrum of the difference between
fractional powers of two self-adjoint operators}, Soviet Mathematics,
{\bf 19}, 1-6 (1975). Translation of Izvestija vyssich.

\bibitem{BrownRavenhall1951}
G.~Brown and D.~Ravenhall,
\newblock \textit{On the interaction of two electrons,}
\newblock {\em Proc. Roy. Soc. London A}, {\bf 208A}, 552--559
(1951).           

\bibitem {BFG} L. Bugliaro,
J. Fr\"ohlich and G.M. Graf, \textit{Stability of
quantum electrodynamics with nonrelativistic matter}, 
Phys.Rev. Lett. {\bf 77} (1996), 3494-3497.


\bibitem{DL}
I.~Daubechies and E.~H. Lieb, \textit{One-electron relativistic molecules
with Coulomb interaction,} Commun. Math. Phys. {\bf 90}, 497-510 (1983).


\bibitem{Evansetal1996} W.~D. Evans, P.~Perry, and H.~Siedentop,
\newblock \textit{The spectrum of relativistic one-electron atoms
according to {B}ethe and {S}alpeter},
\newblock Commun. Math. Phys. {\bf 178}, 733--746 (1996).


\bibitem {FFG} C. Fefferman, J. Fr\"ohlich and G.M. Graf,
\textit{Stability of nonrelativistic quantum mechanical matter coupled to
the (ultraviolet cutoff) radiation field}, Proc. Natl. Acad. Sci. USA
{\bf 93} (1996), 15009-15011; \textit{Stability of ultraviolet cutoff
quantum electrodynamics with non-relativistic matter}, Commun. Math.
Phys. {\bf 190} (1997), 309--330.  


\bibitem{GLL}
M.~Griesemer, E.~H.~Lieb and M.~Loss,
\newblock \textit{Ground states in non-relativistic quantum electrodynamics,}
\newblock Invent. Math. {\bf 145}, 557-595 (2001).

\bibitem{GT}
M.~Griesemer and C. Tix,
\newblock \textit{Instability of a pseudo-relativistic model of matter with
self-generated magnetic field,}
\newblock J.~Math.~Phys. {\bf 40}, 1780-1791 (1999).

\bibitem{HS}
G.~Hoever and H.~Siedentop,
\newblock \textit{Stability of the Brown-Ravenhall operator,}
\newblock  Math.~Phys.~Electronic Jour. {\bf 5}, 1-11 (1999).

\bibitem{Kato}
T. Kato, {\it Perturbation Theory for Linear Operators}, Springer Verlag,
p.307, remark 5.12, 1966.

\bibitem{Anal}
E.H. Lieb and M. Loss,
\newblock {\it Analysis},
\newblock Amer. Math. Soc. second edition, 2001.

\bibitem{selfenergy}
E.~H. Lieb and M.~Loss,
\newblock \textit{Self-energy of electrons in non-perturbative QED,}
\newblock
in  {\it Differential Equations and Mathematical Physics,
University of Alabama, Birmingham, 1999}, R. Weikard and G. Weinstein,
eds. 255-269, Amer. Math. Soc./Internat. Press (2000).  arXiv
math-ph/9908020, mp\_arc 99-305.

\bibitem{selfenergy2}
E.~H. Lieb and M.~Loss,
\newblock \textit{The Ultraviolet problem in on-relativistic QED,}
\newblock  in  preparation.

\bibitem{binding}
E.~H. Lieb and M.~Loss, {\it  A bound on binding energies and mass
renormalization in models of quantum electrodynamics}, Jour. Stat. Phys.
(in press)
arXiv math-ph/0110027.

\bibitem{LLS}
E.~H. Lieb, M.~Loss, and J.~P. Solovej.
\newblock \textit{Stability of matter in magnetic fields,}
\newblock Phys. Rev. Lett. {\bf 75}, 985-989 (1995).

\bibitem {LSS} E.H. Lieb, H. Siedentop and J.P. Solovej,
\textit{Stability and instability of relativistic electrons in magnetic
fields}, J. Stat. Phys. {\bf 89}, 37-59 (1997).


\bibitem{LiebThirring1976}
E.~H. Lieb and W.~E. Thirring,
\newblock \textit{Inequalities for the moments of the eigenvalues of the
Schr\"odinger {H}amiltonian and their relation to {S}obolev inequalities,}
\newblock In E.~H. Lieb, B.~Simon, and A.~S. Wightman, editors, {\em Studies in
Mathematical Physics: Essays in Honor of {V}alentine {B}argmann}. Princeton
University Press, Princeton, 1976.

\bibitem{LY}
E.~H. Lieb and H.-T. Yau.
\newblock \textit{The stability and instability of relativistic matter,}
\newblock Commun. Math. Phys. {\bf 118}, 177--213 (1988).

\bibitem{Loss}
M. Loss
\newblock {\it Stability of matter in magnetic fields}
\newblock In the Proceedings of the XII-th International Congress of
Mathematical Physics 1997, De Wit et al eds., International Press, 1999,
pp. 98-106.

\bibitem{reedsimon1}
M. Reed and B. Simon, {\it Methods of Modern Mathematical Physics}
Vol. I, {\it Functional Analysis}, Academic Press, 1972.

\bibitem{reedsimon2}
M. Reed and B. Simon, {\it Methods of Modern Mathematical Physics}
Vol. II, {\it Fourier Analysis, Selfadjointness}, Academic Press, 1975.

\bibitem{Sucher1980}
J.~Sucher,
\newblock \textit{Foundations of the relativistic theory of many-electron 
atoms,}
\newblock Phys. Rev. A {\bf 22}, 348--362 (1980).

\bibitem{Tix1}
C. Tix,
\newblock \textit{Lower bound for the ground state energy of the no-pair
Hamiltonian,}
\newblock Phys. Lett. B {\bf 405}, 293-296 (1997).

\bibitem{Tix2}
C. Tix,
\newblock \textit{Strict positivity of a relativistic Hamiltonian due to 
Brown and Ravenhall,}
\newblock Bull. London Math. Soc. {\bf 30}, 283-290 (1998).


\end{thebibliography}
\end{document}